\documentclass[prl,aps,twocolumn,superscriptaddress]{revtex4-1}
\usepackage{graphicx,color}
\usepackage{amsthm}
\usepackage{amsfonts}
\usepackage{algorithmic}
\usepackage{enumerate}
\usepackage{latexsym}
\usepackage{amsmath}
\usepackage{amssymb}
\usepackage{multirow}
\usepackage{array}
\usepackage{diagbox}
\usepackage{subfigure}
\usepackage[colorlinks=true,citecolor=blue,linkcolor=blue]{hyperref}
\def\avg#1{\left\langle#1\right\rangle}

\emergencystretch=\maxdimen
\begin{document}

\title{Antiferromagnetic fluctuations and dominant $d_{xy}$-wave pairing symmetry in nickelate-based superconductors}
\author{Chao Chen}
\affiliation{Department of Physics, Beijing Normal University, Beijing 100875, China\\}
\author{Runyu Ma}
\affiliation{Department of Physics, Beijing Normal University, Beijing 100875, China\\}
\author{XueLei Sui}
\affiliation{Beijing Computational Science Research Center, Beijing 100084, China}
\author{Ying Liang}
\affiliation{Department of Physics, Beijing Normal University, Beijing 100875, China\\}
\author{Bing Huang}
\affiliation{Beijing Computational Science Research Center,
Beijing 100084, China}
\affiliation{Department of Physics, Beijing Normal University, Beijing 100875, China\\}
\author{Tianxing Ma}
\email{txma@bnu.edu.cn}
\affiliation{Department of Physics, Beijing Normal University, Beijing 100875, China\\}
\affiliation{Beijing Computational Science Research Center,
Beijing 100084, China}

\begin{abstract}
Motivated by recent experimental studies on superconductivity found in nickelate-based materials, we study the temperature dependence of the spin correlation and the superconducting pairing interaction within an effective two-band Hubbard model by the quantum Monte Carlo method.
Based on parameters extracted from first-principles calculations, our intensive numerical results reveal that the pairing with a $d_{xy}$-wave symmetry firmly dominates over other pairings at low temperature, which is mainly determined by the Ni 3$d$ orbital. It is also found that the effective pairing interaction is enhanced as the on-site interaction increases, demonstrating that the superconductivity is driven by strong electron-electron correlation. Even though the $(\pi,\pi)$ antiferromagnetic correlation could be enhanced by electronic interaction, there is no evidence for long-range antiferromagnetic order exhibited in nickelate-based superconductors. Moreover, our results offer possible evidence that the pure electron correlation may not account for the charge density wave state observed in nickelates.

\end{abstract}
\maketitle

\noindent
\section{I. Introduction}
Understanding the mechanism of high-Tc superconductivity \cite{PhysRev.108.1175,Bednorz1986,doi:10.1126/science.235.4793.1196,RevModPhys.60.585,Anderson_2004,RevModPhys.78.17,RevModPhys.84.1383} and intertwining symmetry-breaking orders \cite{RevModPhys.87.457,annurev-conmatphys,Keimer2015} has always been a central issue in condensed matter physics. Recently, the discovery of superconductivity in the family of Sr-doped RNiO$_2$ (R=Nd, La, Pr) \cite{Li2019,SawatzkyAUG29,PhysRevLett.125.147003,arXiv:2201.10038,PhysRevLett.125.027001} has attracted great research interest, which may provide a new opportunity for further understanding unconventional superconductivity \cite{PhysRevB.101.020503,PhysRevLett.124.166402,PhysRevB.101.064513,PhysRevB.102.224506,PhysRevMaterials.4.121801,PhysRevB.100.201106,PhysRevMaterials.4.084409,LeeAPR1,OsadaNOV,zeng2021superconductivity}.
Among them, one essential object is to identify the dominant superconducting pairing form, which remains a major challenge of today's studies on this family. In a single-particle tunneling experiment on a Sr-doped NdNiO$_2$ film surface, researchers detected singlet pairing, but they could not distinguish whether it is an $s$ wave, $d$ wave or their mixture \cite{Gu2020}.
At present, some theoretical studies of nickelate-based superconductors have been based on models with one-orbital (Ni 3$d$) band structures that support these materials being captured by a one-band Hubbard model \cite{Zhang_2021,PhysRevResearch.2.023112,Kitatani2020}, and they have revealed a dominant $d$-wave pairing in their model \cite{Zhang_2021,PhysRevResearch.2.023112}. However, others have proposed various possibilities for multiband models \cite{PhysRevB.102.100501,PhysRevB.101.060504,PhysRevB.101.020501,lu2021twoorbital,PhysRevLett.125.077003,PhysRevLett.129.077002}. The $t-J-K$ model, which considers the Kondo coupling, exhibits a transition between the $d$ wave and ($d+is$) wave of the dominant pairing at large hole doping \cite{PhysRevB.102.220501}. Research on the controversial pair symmetry of nickelates is necessary both experimentally and theoretically.
From the theoretical viewpoint, using unbiased numerical techniques is believed to be the only opportunity to achieve this goal if the electronic correlation dominates in the system.

Besides the superconductivity, the spin density wave (SDW) \cite{PhysRevResearch.4.023093,LuJUL9,Cui_2021} and charge density wave (CDW) \cite{Rossi2022,Tam2022,PhysRevLett.129.027002,peng2021superconductivity,arXiv:2204.12208,arXiv:2207.00266}, which are observed in nickelates, also attract high attention in quest of their origins. Previous theoretical works on density functional theory (DFT) \cite{PhysRevB.59.7901,PhysRevB.70.165109,PhysRevX.10.011024,PhysRevX.11.011050,Hepting2020} have systematically studied the characteristics of nickelate electronic structures. It is found that there are both similarities and differences compared with those of cuprates \cite{lu2021twoorbital,PhysRevLett.124.207004,PhysRevB.100.205138,PhysRevResearch.1.032046,PhysRevX.10.011024,PhysRevX.11.011050,Hepting2020,PhysRevB.101.060504}. These results provide a cornerstone to study the magnetism, superconductivity and CDW in the nickelate family. According to the DFT calculation of RNiO$_2$ \cite{PhysRevX.10.011024,PhysRevX.11.011050,Hepting2020}, the two bands near $E_F$ mainly contributed to its physical properties. One band, composed of Ni 3$d_{x^{2}-y^{2}}$ and O 2$p$ orbitals, has a Zhang-Rice-singlet-like character, while the contribution of oxygen in the nickelates is smaller than that in cuprates, and the other band, composed of the R 5$d$ orbital, forms an important metallic electron pocket. These two orbitals hybridize, forming a two-band system, where the strongly correlated Ni layers play an important role \cite{PhysRevX.10.011024,PhysRevX.11.011050,Hepting2020}.

\begin{figure}[h]
\includegraphics[scale=0.56, trim = 100 380 100 290, clip]{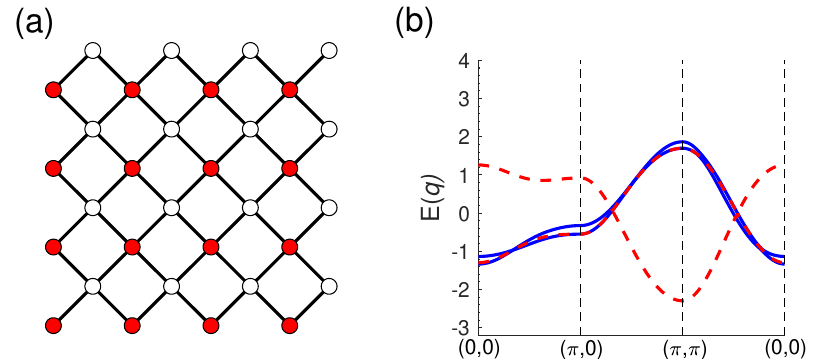}
%zuo, xia ,you,shang
\caption{(Color online)(a) Here, red and white circles indicate different sublattices, A and B. The nearest distance between B and B (or A and A) is 2.
(b) The energy band along the high-symmetry line in the unfolded Brillouin zone. Solid blue lines: $k_z$=0; dashed red lines: $k_z={\pi}$ in Table \ref{table1}. }
\label{Fig:Sketch}
\end{figure}
%zuo, xia ,you,shang

To identify the superconducting pairing form of nickelate-based materials, we perform a quantum Monte Carlo study of the spin correlation and superconducting pairing interaction in an effective two-band microscopic model
based on  parameters extracted from first-principles calculations.
From the results of the Wannier orbitals \cite{Hepting2020,PhysRevX.10.011024}, a two-band model is constructed that contains two main bands near $E_F$, and this model also contains inter-orbital coupling between the Ni 3$d$ orbital and the R $5d$ orbital. The calculations of the pairing correlation show that there exists an extensive $d$-wave channel that firmly dominates over other pairings at low temperature and the pairing channel is determined by the Ni 3$d$ orbital.
For different fillings $\avg{n}$=1.0, 0.9, and 0.8, the $(\pi,\pi)$ antiferromagnetic (AFM) correlation and the effective pairing interaction are both enhanced as the on-site interaction increases.
Our unbiased calculations demonstrate that the superconductivity and AFM correlation in nickelate-based superconductors should be driven by electron-electron correlation.
Although the $(\pi,\pi)$ antiferromagnetic correlation could be enhanced by electronic interaction, there is no evidence for long-range antiferromagnetic order exhibited in nickelate-based superconductors.
Additionally, by considering the nearest-neighbor repulsion of the Ni 3$d$ orbital, the CDW state exhibits a $q=(\pi,\pi)$ pattern.
%This conclusion support many proposed models\cite{Zhang_2021,lu2021twoorbital}, and give an exact numerical calculation result.

%A tuning metal-insulator transition and superconductivity in TBG is proposed by studying
%the interlay coupling strength dependent spin correlation, conductivity and superconducting pairing interactions.

\noindent
\section{II. Model and methods}
In the two-band Hubbard model, the tight-binding part contains intralayer hopping, interlayer hopping and the strongly correlated Ni layer. Therefore, the nickel-square lattice Hamiltonian can be written as
\begin{eqnarray}
H=&&H_{1}+H_{2}+H_{3}+H_{4}, \notag \\
H_{1}=&&t_3^{Nd-Ni}\sum_{\mathbf{i}\eta \sigma }[a_{\mathbf{i}\sigma }^{\dag }b_{%
\mathbf{i}+\eta\sigma }+h.c.],  \notag \\
H_{2}=&&t_1^{Nd}[\sum_{\mathbf{i}\tau_{1}\sigma }a_{\mathbf{i}\sigma }^{\dag }a_{\mathbf{i}%
+\tau_{1}\sigma }]+t_2^{Nd}[\sum_{\mathbf{i}\tau_{2}\sigma }a_{\mathbf{i}\sigma }^{\dag }a_{\mathbf{i}+\tau_{2}\sigma }]\notag \\
+&&t_3^{Nd}[\sum_{\mathbf{i}\tau_{3}\sigma }a_{\mathbf{i}\sigma }^{\dag }a_{\mathbf{i}+\tau_{3}\sigma }],\notag \\
H_{3}=&&t_1^{Ni}[\sum_{\mathbf{i}\tau_{1}\sigma }b_{\mathbf{i}\sigma }^{\dag }b_{\mathbf{i}%
+\tau_{1}\sigma }]+t_2^{Ni}[\sum_{\mathbf{i}\tau_{2}\sigma }b_{\mathbf{i}\sigma }^{\dag }b_{\mathbf{i}+\tau_{2}\sigma }]\notag \\
+&&t_3^{Ni}[\sum_{\mathbf{i}\tau_{3}\sigma }b_{\mathbf{i}\sigma }^{\dag }b_{\mathbf{i}+\tau_{3}\sigma }],\notag \\
H_{4}=&&U\sum_{\mathbf{i}}n_{b\mathbf{i}\uparrow}n_{b\mathbf{i}\downarrow}
+\mu\sum_{\mathbf{i}\sigma}[(1+\Delta/\mu)n_{a\mathbf{i}\sigma }+n_{b\mathbf{i}\sigma}]
\end{eqnarray}
Here, $a_{\mathbf{i}\sigma}$ ($a_{\mathbf{i}\sigma}^{\dag}$) annihilates (creates) electrons
at site $\mathbf{R}_\mathbf{i}$ with spin $\sigma$ ($\sigma$=$\uparrow,\downarrow$)
on sublattice A, $b_{\mathbf{i}\sigma}$ ($b_{\mathbf{i}\sigma}^{\dag}$) annihilates (creates)
electrons at site $\mathbf{R}_\mathbf{i}$ with spin $\sigma$
($\sigma$=$\uparrow,\downarrow$) on sublattice B,
$n_{a\mathbf{i}\sigma}=a_{\mathbf{i}\sigma}^{\dagger}a_{\mathbf{i}\sigma}$,
$n_{b\mathbf{i}\sigma}=b_{\mathbf{i}\sigma}^{\dagger}b_{\mathbf{i}\sigma}$, $\eta=(\pm 3\hat{x} ,\pm3\hat{y})$, $\tau_1=(\pm 2\hat{x},0)$ and $(0,\pm 2\hat{y})$, $\tau_2=(\pm 2\hat{x} ,\pm2\hat{y})$, and $\tau_3=(\pm 4\hat{x},0)$ and $(0,\pm 4\hat{y})$.
%$t_1$ is the NN hopping integral and $\mu$ the chemical potential.
Our first principles calculations give consistent on-site energy and hopping parameters with previous works. \cite{Hepting2020,PhysRevX.10.011024,PhysRevX.11.011050}. For more details about our Wannier downfolding of NdNiO$_2$, please see Table \ref{table2} in the Appendix 1. For simplicity and clarity, we mainly take the parameters from Refs. \cite{Hepting2020,PhysRevX.10.011024,PhysRevX.11.011050} and list the hopping parameters of NdNiO$_2$ that we use in Table \ref{table1} at $k_z$=0, $\pi$$/2$ and $\pi$.
From the analysis of the first-principles calculations \cite{PhysRevX.10.011024,Hepting2020,PhysRevX.10.021061,PhysRevX.10.041002,PhysRevX.11.011050}, $\Delta$=$\Delta_1$-$\Delta_2$ represents the on-site energy difference between the Nd 5$d$ orbital and the Ni 3$d$ orbital. In the following calculations, we mainly discuss the cases of $k_z$=0 and $k_z=\pi$.

\begin{table}[tbp]
    \begin{tabular}{p{1cm}<{\centering} p{2cm}<{\centering} p{2cm}<{\centering} p{2cm}<{\centering} }
        \hline\hline
        \noalign{\smallskip}
        \multicolumn{4}{c}{Hopping parameters for the tight binding model}\\
        \noalign{\smallskip}
        \hline
        \diagbox{$t$}{$k_z$} & $0$ & $\pi / 2$ & $\pi$ \\
        \hline
        \noalign{\smallskip}
        \multicolumn{4}{c}{$t^{Nd}$}\\
        \noalign{\smallskip}
        $\Delta_{1}$ & 0.633 & 1.305 & 1.287 \\
        $t_{1}$ & -0.380 & -0.028 & 0.444 \\
        $t_{2}$ & 0.084 & -0.090 & -0.180 \\
        $t_{3}$ & 0.003 & 0.027 & 0.051 \\
        \noalign{\smallskip}
        \multicolumn{4}{c}{$t^{Ni}$}\\
        \noalign{\smallskip}
        $\Delta_{2}$ & 0.242 & 0.308 & 0.374 \\
        $t_{1}$ & -0.374 & -0.374 & -0.374 \\
        $t_{2}$ & 0.094 & 0.094 & 0.094 \\
        $t_{3}$ & -0.043 & -0.043 & -0.043 \\
        \noalign{\smallskip}
        \multicolumn{4}{c}{$t^{Nd-Ni}$}\\
        \noalign{\smallskip}
        $t_{3}$ & 0.020 & 0.020 & 0.020 \\
        \hline\hline
    \end{tabular}
    \caption{Hopping parameters (in units of eV) for the tight binding model from Refs.\cite{Hepting2020,PhysRevX.10.011024,PhysRevX.11.011050}.}
    \label{table1}
\end{table}

Our simulations are mainly performed on the lattice shown in Fig.~\ref{Fig:Sketch}(a) of $L$=8 (the total number of lattice sites is $N_s$=2$\times$$L^2$=128) by using the determinant quantum Monte Carlo (DQMC) method at finite temperature with periodic boundary conditions.
The basic strategy of the DQMC method is to express the partition function as high-dimensional integrals on a set of random auxiliary fields. Then, the Monte Carlo techniques complete the integral.
In the simulations, we use 3000 sweeps to equilibrate the system and an additional 10000--40000 sweeps to generate measurements. These measurements were split into 10 bins and provided the basis of coarse-grain averages. The errors were calculated based on the standard deviation from the average. For more technical details, please see Refs.\cite{PhysRevD.24.2278,doi:10.1063/1.3485059,PhysRevLett.110.107002,PhysRevB.40.506}, as well as information in the Appendix.

As magnetic excitation possibly plays a significant role in the superconductivity mechanism of electronic correlated systems, we investigate the spin susceptibility in the $z$ direction at zero frequency,
\begin{eqnarray}
\chi(q)=\int_{0}^{\beta}d\tau \sum_{d,d'=a,b} \sum_{\mathbf{i,j}}
e^{iq\cdot(\mathbf{i}_{d}-\mathbf{j}_{d'})} \langle\textrm{m}_{\mathbf{i}_{d}}(\tau) \cdot
\textrm{m}_{\mathbf{j}_{d'}}(0)\rangle,\notag \\
\end{eqnarray}
where $m_{\mathbf{i}_{a}}(\tau)$=$e^{H\tau}m_{\mathbf{i}_{a}}(0)e^{-H\tau}$ with
$m_{\mathbf{i}_{a}}$=$a^{\dag}_{\mathbf{i}\uparrow}a_{\mathbf{i}\uparrow}-a^{\dag}_{\mathbf{i}\downarrow}a_{\mathbf{i}\downarrow}$
and $m_{\mathbf{i}_{b}}$=$b^{\dag}_{\mathbf{i}\uparrow}b_{\mathbf{i}\uparrow}-b^{\dag}_{\mathbf{i}\downarrow}b_{\mathbf{i}\downarrow}$.
To study the superconducting property of nickelate-based superconductors, we calculated the pairing susceptibility,
\begin{equation}
P_{\alpha }=\frac{1}{N_{s}}\sum_{\mathbf{i,j}}\int_{0}^{\beta }d\tau \langle \Delta
_{\alpha }^{\dagger }(\mathbf{i},\tau )\Delta_{\alpha }^{\phantom{\dagger}%
}(\mathbf{j},0)\rangle ,
\label{shi3}
\end{equation}
where $\alpha $ denotes the pairing symmetry. Due to the constraint of different
on-site Hubbard interaction in two sublattices, pairing between the same sublattices is
favored, and the corresponding order parameter $\Delta_{\alpha }^{\dagger
}(\mathbf{i})$\ is written as
\begin{equation*}
\Delta _{\alpha }^{\dagger }(\mathbf{i})\ =\sum_{l}f_{\alpha }^{\dagger }(\delta_{l})(a_{{\mathbf{i}}\uparrow }b_{{\mathbf{i}+{\delta_{l}}}\downarrow }-a_{{\mathbf{i}}\downarrow }b_{{%
\mathbf{i}+\delta_{l}}\uparrow })^{\dagger },
\end{equation*}%
where $f_{\alpha}(\mathbf{{\delta_l})}$ stands for the form factor of the pairing
function. The vectors $\mathbf{{\delta_l}}$ ($l$=1,2,3,4) represent
the nearest intersublattice connections, where $\mathbf{\delta}$ is $(\pm \hat{x},\pm \hat{y})$, or the nearest intrasublattice connections
where $\mathbf{\delta'}$ is $(\pm 2\hat{x},0)$ and $(0,\pm 2\hat{y})$.

Furthermore, in order to explore the CDW state, we define the density-density correlation function\cite{Wang_2014,PhysRevB.91.241117,PhysRevLett.122.077602},
\begin{equation}
C(R)=\frac{1}{N_{s}N_{R}}\sum_{\mathbf{i}}\sum_{\mathbf{|j-i|=R}}
\langle(n_{i}-\langle n_{i} \rangle)(n_{j}-\langle n_{j} \rangle)\rangle,
\label{shi4}
\end{equation}
Here, $n_{i}$ and $n_{j}$ denote the electronic number operator at site $\mathbf{R}_\mathbf{i}$ and $\mathbf{R}_\mathbf{j}$. $R$ is the distance between site i and site j. The $N_R$ is the total number of distance $R$. And its Fourier transform can be written as,
\begin{equation}
C(q)=\frac{1}{N_{s}}\sum_{\mathbf{R}}e^{iqR}C(R),
\label{shi5}
\end{equation}

\noindent
\section{III. Results and discussion}
\begin{figure}[tbp]
\includegraphics[scale=0.29,trim = 175 100 10 50, clip]{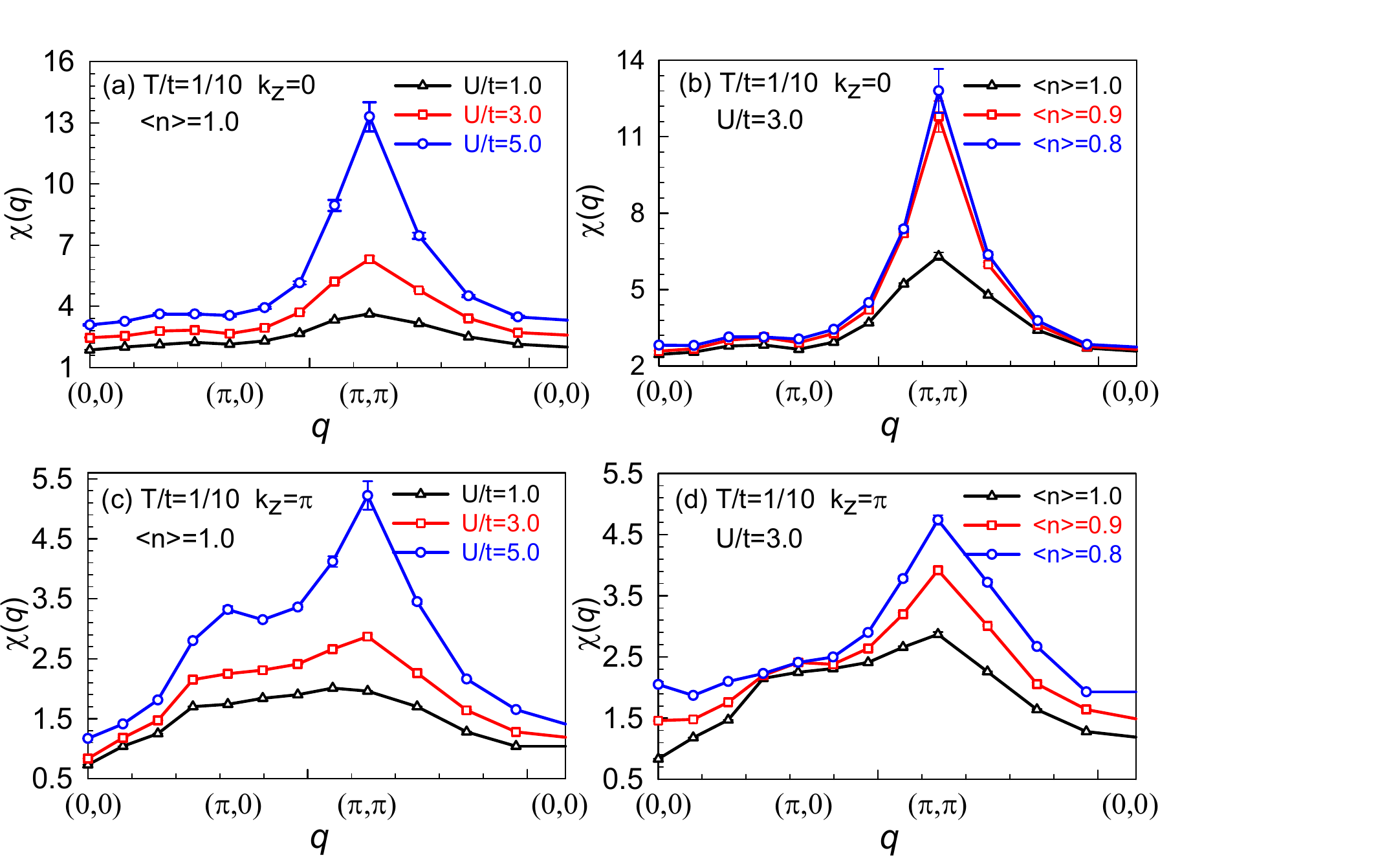}
\caption{(Color online) Magnetic susceptibility $\chi(q)$ versus
momentum $q$, (a) for different $U$ at $\avg{n}=1.0$, (b) for different fillings at $U/t=3.0$ (where $t=|t^{Ni}_{1}|=0.374eV$) and $k_z=0$ on a $2\times8^2$ lattice;
(c) for different $U$ at $\avg{n}=1.0$, (d) for different fillings at $U/t=3.0$ and $k_z=\pi$ on a $2\times8^2$ lattice.}
\label{Fig:spin}
\end{figure}

To study the magnetic correlations, we calculated the spin susceptibility $\chi(\textbf{q})$ in Fig.~\ref{Fig:spin} at different $U$ and fillings $\avg{n}$ at temperature $T/t$=1/10. In Fig.~\ref{Fig:spin}, one can notice that there is a sharp peak at $(\pi,\pi)$, which indicates the domination of AFM correlation at both $k_z=0$ and $k_z=\pi$.
In Fig.~\ref{Fig:spin} (a) and Fig.~\ref{Fig:spin}(c), we can see that the AFM correlation is enhanced as $U$ increases, which indicates that such an AFM correlation is driven by strong electron-electron correlation.
Fig.~\ref{Fig:spin} (b) and Fig.~\ref{Fig:spin} (d) shows that the peak is enhanced at fillings $\avg{n}$=0.9 and 0.8, which indicates that the AFM correlation is promoted when the system is doped away from half filling.
Recently, resonant inelastic x-ray scattering experiments have revealed an AFM exchange interaction\cite{LuJUL9}.
Our results here might provide evidence for the AFM exchange couplings observed in infinite-layer nickelates.

%zuo, xia ,you,shang

\begin{figure}[tbp]
\includegraphics[scale=0.29,trim = 185 40 170 10, clip]{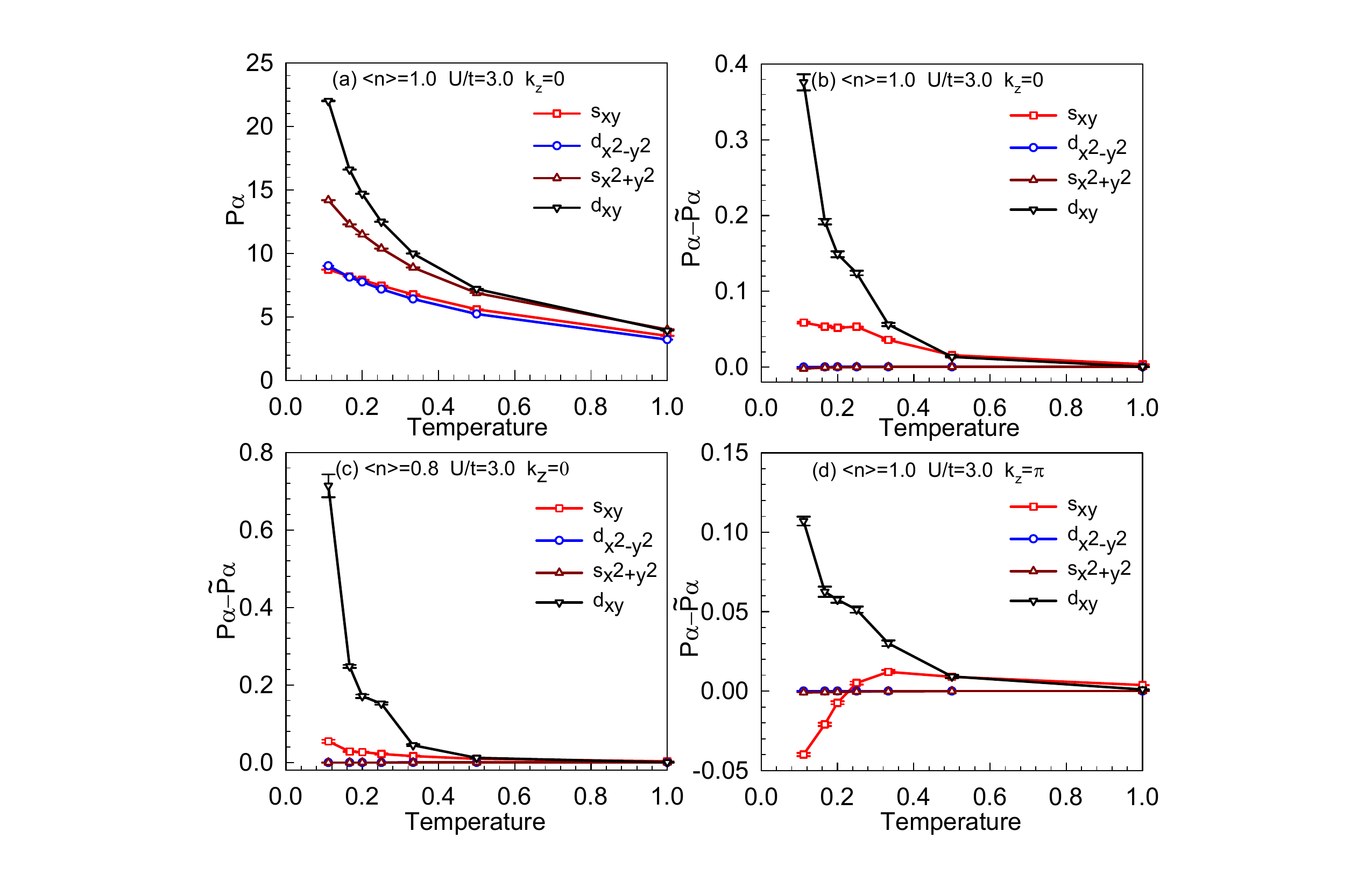}
\caption{(Color online) (a) Pairing susceptibility $P_{\alpha}$ and (b) the effective pairing interaction $P_{\alpha}-\tilde{P}_{\alpha}$ as a function
of temperature for different pairing symmetries at $\avg{n}=1.0$, $U/t=3.0$ and $k_z=0$ on a $2\times8^2$ lattice.
(c) The effective pairing interaction $P_{\alpha}-\tilde{P}_{\alpha}$ as a function of temperature for different pairing symmetries at $\avg{n}=0.8$, $U/t=3.0$ and $k_z=0$ on a $2\times8^2$ lattice, (d) at $\avg{n}=1.0$, $U/t=3.0$ and $k_z=\pi$ on a $2\times8^2$ lattice.}
\label{Fig:FigP}
\end{figure}

In Fig.~\ref{Fig:FigP} (a), we show the temperature dependence of the pairing
susceptibilities for different pairing symmetries at half filling with $U/t=3.0$ at $k_z=0$.
We can clearly observe that the pairing susceptibilities for various pairing symmetries increase with decreasing temperature. Most strikingly, $d_{xy}$ increases
much faster than any other pairing symmetry as the temperature is lowered. This indicates that the $d_{xy}$ pairing symmetry is dominant over the other pairing symmetry at half filling.
Our further results also illustrate that the $d_{xy}$ pairing symmetry is robust at different fillings and $U$.

%zuo, xia ,you,shang
The effective pairing interaction is a direct probe for the superconductivity.
To extract the effective pairing interaction, the uncorrelated single-particle contribution $\widetilde{P}
_{\alpha }(\mathbf{i,j})$ is calculated, which is achieved by replacing $\langle
a_{\mathbf{i}\downarrow }^{\dag }b_{\mathbf{j}\uparrow }a_{\mathbf{i}+\delta_{l}\downarrow}^{\dag}
b_{\mathbf{j}+\delta_{l'}\uparrow}\rangle $ in Eq.~\ref{shi3} with $\langle a_{\mathbf{i}\downarrow }^{\dag
}b_{\mathbf{j}\uparrow }\rangle \langle a_{\mathbf{i}+\delta_{l}\downarrow }^{\dag }
b_{\mathbf{j}+\delta_{l'}\uparrow }\rangle $, and then we get the effective pairing interaction $P_{\alpha}-\widetilde{P}_{\alpha}$.
In Fig.~\ref{Fig:FigP} (b) and Fig.~\ref{Fig:FigP} (c), it is obvious that $P_{\alpha}-\widetilde{P}_{\alpha}$ presents a very similar temperature dependence to that of $P_{\alpha}$ at $\avg{n}=1.0$ or $\avg{n}=0.8$.
Moreover, the effective pairing interaction for $d_{xy}$ pairing is always positive and increases much faster than any other pairing symmetry at low temperatures. Such a temperature dependence shows that there indeed exists attraction for the $d_{xy}$ pairing at $k_z=0$.
From Fig.~\ref{Fig:FigP} (d), we can find that the $d_{xy}$ pairing symmetry is also dominant at $k_z=\pi$. Therefore, although hopping parameters $t_i^{Nd}$ and  the on-site energy difference $\Delta$ are changed at different $k_z$, the investigated magnetism and pairing interaction show identical physical results. In the following, we mainly discuss hopping parameters at $k_z=0$ for simplicity.

\begin{figure}[tbp]
\includegraphics[scale=0.28,trim = 150 280 140 145, clip]{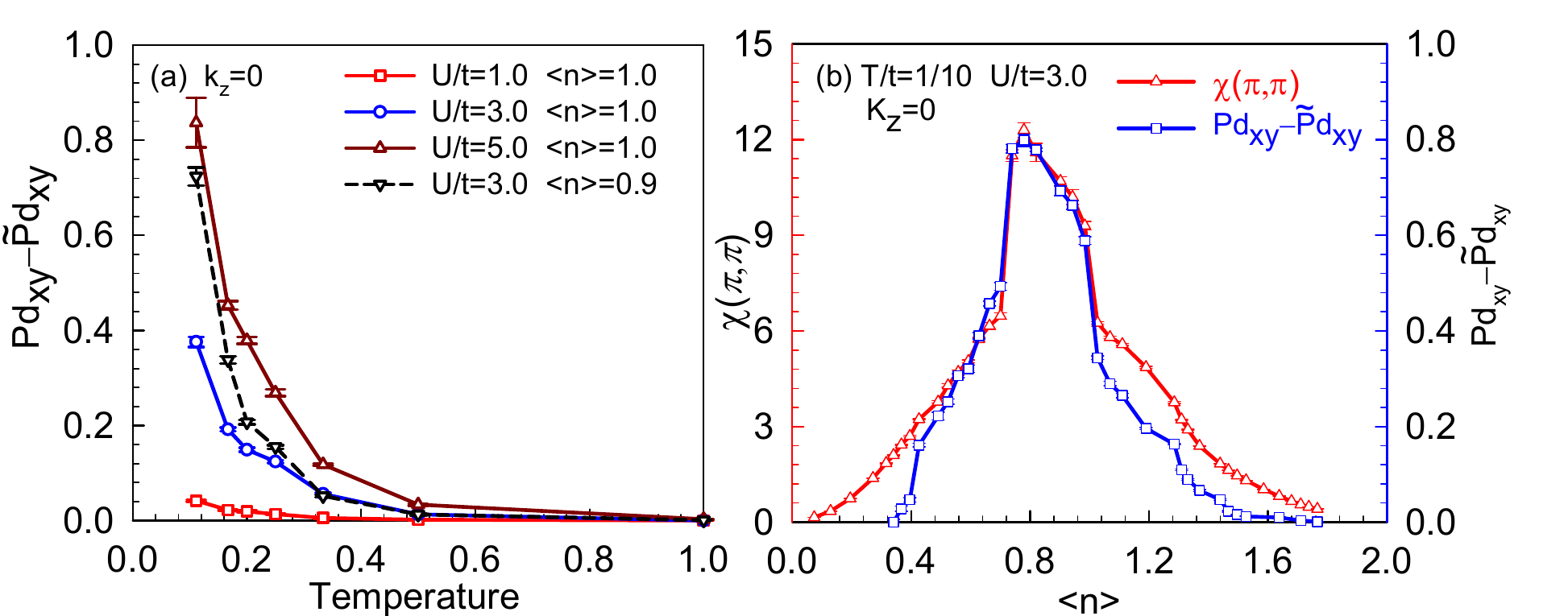}
\caption{(Color online) (a) The effective pairing interaction $P_{d_{xy}}-\tilde{P}_{d_{xy}}$ as a function of temperature for different $U$ at $\avg{n}=1.0$ and $k_z=0$ on a $2\times8^2$ lattice. (b) The effective pairing interaction $P_{d_{xy}}-\tilde{P}_{d_{xy}}$ and the $(\pi,\pi)$ AFM correlation $\chi(\pi,\pi)$ as a function of fillings at $T/t=1/10$, $U/t=3.0$ and $k_z=0$ on a $2\times8^2$ lattice. }
\label{Fig:Fign}
\end{figure}

Fig.~\ref{Fig:Fign} (a) shows the effective pairing interaction as
a function of temperature for the $d_{xy}$ wave at different $U$. We can see that the effective pairing interaction of the $d_{xy}$ wave is enhanced with increasing $U$. For $U/t=1.0$, the effective pairing interaction $P_{d_{xy}}-\tilde{P}_{d_{xy}}$ is very small even in the low-temperature region, which may be due to the small AFM structure of the system in Fig.~\ref{Fig:spin}(a).
For $U/t=3.0$ and $U/t=5.0$, remarkably, the effective pairing interaction $P_{d_{xy}}-\tilde{P}_{d_{xy}}$ tends to diverge at low temperatures, and with increasing $U$, this divergence tends to be promoted. This indicates that the $d_{xy}$ pairing superconductivity should be driven by a strong electron-electron correlation.

%zuo, xia ,you,shang        trim = 0 250 50 50,
In Fig.~\ref{Fig:Fign} (b), we studied the filling dependence of the effective pairing interaction $P_{d_{xy}}-\tilde{P}_{d_{xy}}$ and the $(\pi,\pi)$ AFM correlation $\chi(\pi,\pi)$ at $T/t=1/10$, $U/t=3.0$ and $k_z=0$. Fig.~\ref{Fig:Fign} (b) indicates that the optimal electron filling is slightly below $\avg{n}=0.8$, where the effective pairing interaction and the AFM correlation is largest.
Fig.~\ref{Fig:spin} and Fig.~\ref{Fig:Fign} show that the increase in the peak at ($\pi, \pi$) of spin susceptibility is correlated with the promotion of the pairing susceptibility. This directly confirms that the $(\pi,\pi)$ AFM fluctuations enhance the $d_{xy}$ pairing.

From the above studies, we know that the system exhibits local antiferromagnetism. To further explore whether there is a long-range AFM order, we also calculate the AFM spin structure factor,
\begin{equation}
S_{AFM}=\frac{1}{N_s}\langle [\sum_{{r}}(-1)^r{\hat{S}^{z}_{br}}]^{2}  \rangle,
\label{spin}
\end{equation}
Here, ${\hat{S}^{z}_{br}}$ is the $z$ component spin operator on the B sublattice. When $\lim_{N_s\rightarrow\infty}(S_{AFM}/N_s)>$0, it indicates the onset of long-range AFM order.
\begin{figure}[tbp]
\includegraphics[scale=0.3,trim = 140 40 220 80, clip]{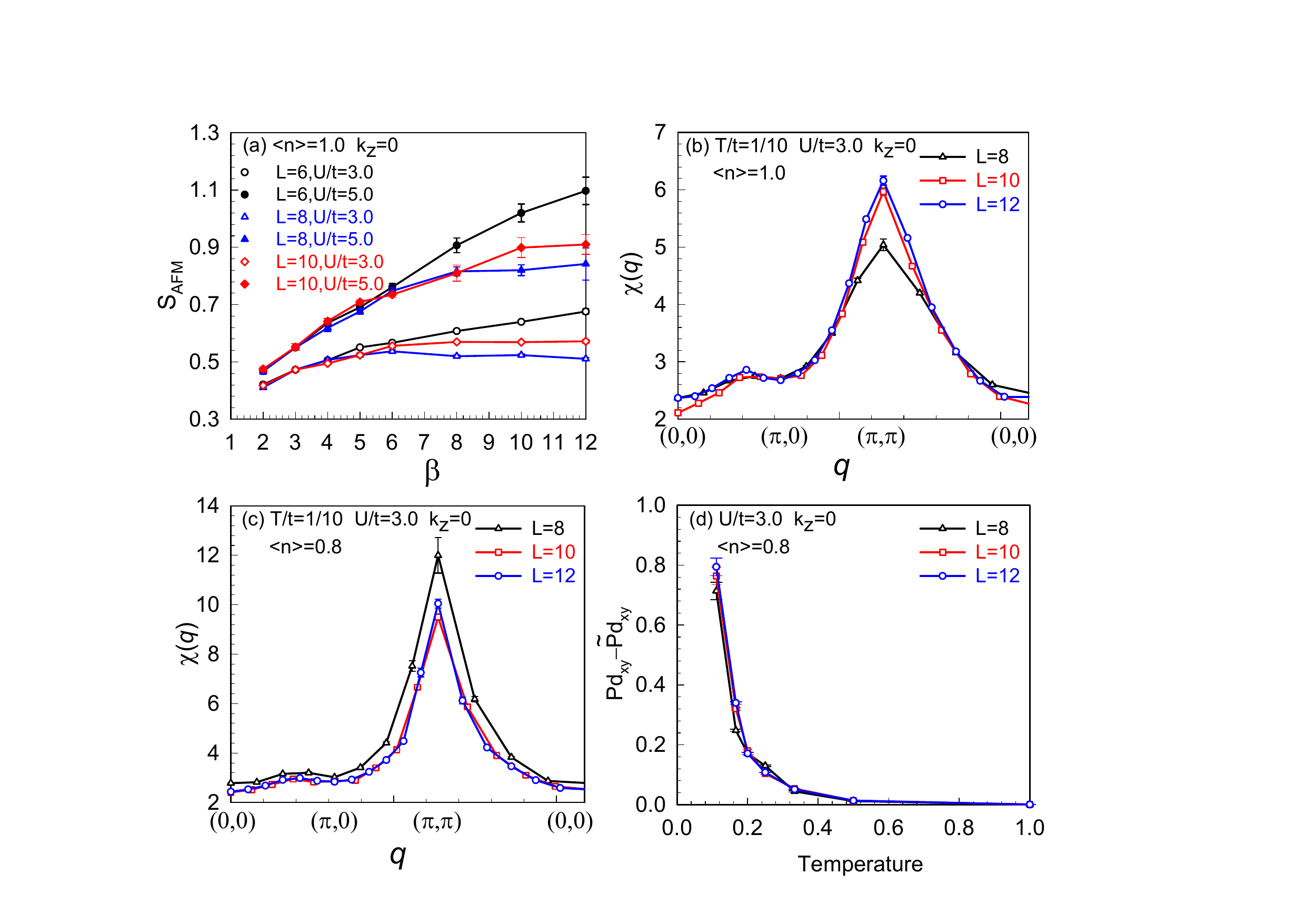}
\caption{(Color online) (a) The AFM spin structure factor $S_{AFM}$ depends on $\beta=1/T$ with different interaction strengths and lattice sizes at $\avg{n}=1.0$ and $k_z=0$. 
(b) Magnetic susceptibility $\chi(q)$ versus momentum $q$ for different lattice sizes at $U/t=3.0$, $T/t=1/10$, $k_z=0$ for $\avg{n}=1.0$  and (c) $\avg{n}=0.8$. (d) The effective 
pairing interaction $P_{d_{xy}}-\tilde{P}_{d_{xy}}$ as a function of temperature for different lattice sizes at $\avg{n}=0.8$, $U/t=3.0$ and $k_z=0$.}
\label{Fig:Figafm}
\end{figure}
%zuo, xia ,you,shang
In Fig.~\ref{Fig:Figafm} (a), we present the results of the AFM spin structure factors as a function of $\beta$ for different interaction strengths $U$ and lattice sizes $L$, which demonstrates the spin structure factor is nearly saturated at $\beta$=10.
Interesting, $S_{AFM}$ decreases as the lattice size increases at low temperatures,
which indicates that there is no long-range AFM order at $U/t\le5.0$ and $\avg{n}=1.0$.
In Fig.~\ref{Fig:Figafm} (b) and Fig.~\ref{Fig:Figafm} (c),
it is shown that $\chi(q)$ has a very minor size dependency with lattice sizes $L = $10 and 12 at $\avg{n}=1.0$ or $\avg{n}=0.8$.
Actually, it is more difficult to exhibit long-range AFM order at $\avg{n}=0.8$,
since $\chi(q)$ decreases as the lattice size increases from $L$=8 to 10.
Different from the AFM spin structure factors, as that shown in Fig.~\ref{Fig:Figafm} (d),
the effective pairing interaction increases very fast as the temperature decreases,
and has a potential to diverge as the temperature is low enough.
Moreover, $P_{d_{xy}}-\tilde{P}_{d_{xy}}$ increases slightly as the system size increases.
These two facts, different from the magnetic order, indicate that the superconducting order with $d_{xy}$ symmetry
should survive even at thermodynamic limit. Therefore, our numerical results reveal that the $d_{xy}$-wave symmetry firmly
dominates over other pairings and the system may exhibit superconductivity as the temperature is low enough.
Two closely related theoretical works by DMFT also report the absence of long-range AFM order and its competition with superconductivity \cite{PhysRevB.105.205131,fphy.2022.834682}.

At last, to discuss the electron correlation effect on the CDW state, we consider the nearest-neighbor repulsion of the Ni 3$d$ orbital in the Hamiltonian, which can be written as
\begin{eqnarray}
H_{V}=&&V\sum_{\mathbf{i,\tau_{1}}}n_{b\mathbf{i}}n_{b\mathbf{(i+\tau_{1})}},
\end{eqnarray}
\begin{figure}[tbp]
\includegraphics[scale=0.295,trim = 192 450 140 30, clip]{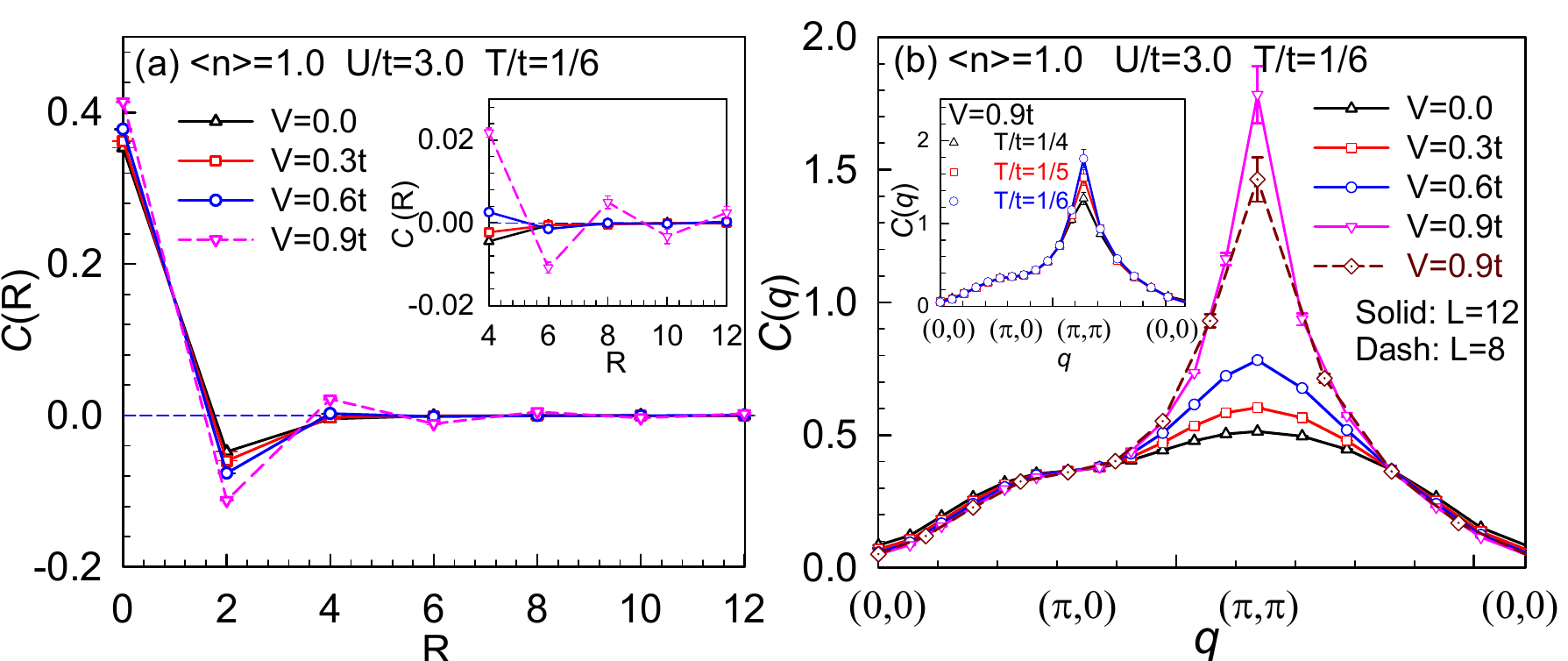}
\caption{(Color online) (a) The density-density correlations $C(R)$ of Ni 3$d$ orbital as a function of distance $R$ for different $V$ at $\avg{n}=1.0$, $U/t=3.0$ and temperature $T/t=1/6$ on a $2\times12^2$ lattice. Inset: The enlarged $C(R)$ for $R\ge4.0$. (b) The density-density correlations $C(q)$ versus momentum $q$ for different $V$ at $\avg{n}=1.0$, $U/t=3.0$ and $T/t=1/6$ on a $2\times12^2$ or $2\times8^2$ lattice. Inset: $C(q)$ versus momentum $q$ for different $T$ at $\avg{n}=1.0$, $U/t=3.0$ and $V/t=0.9$ on a $2\times12^2$ lattice.
  }
\label{Fig:Figcdw}
\end{figure}
%zuo, xia ,you,shang

In Fig.~\ref{Fig:Figcdw} (a), we can notice the density-density correlation  function $C(R)$ develops a staggered pattern as the interaction strength increases to $V=0.9t$, which indicates the onset of the CDW.
Fig.~\ref{Fig:Figcdw} (b) shows that the peak of $q=(\pi,\pi)$ is quickly enhanced at $V=0.9t$, which also is a signal of the CDW's presence. Due to the serious sign problem at low temperature or high interaction, we only display the temperature effect at $T/t=1/4, 1/5, 1/6$ and $V=0.9t$ in the inset of Fig.~\ref{Fig:Figcdw} (b) and can see a small enhancement of charge correlations with decreasing temperature.

\noindent
\section{IV. Summary}
In summary, within an effective two-band model for nickelate-based superconductors, we study the spin correlation, the superconducting pairing interaction, and the density-density correlation by using the unbiased numerical techniques of DQMC.
We identify that the $d_{xy}$ wave pairing channel is dominant in nickelate-based superconductors, which might support
the recent London penetration depth experiment \cite{arxiv.2201.12971}. Both the $(\pi,\pi)$ AFM and
the pairings with the $d_{xy}$ symmetry are enhanced with increasing
electron-electron correlation, especially in the low-temperature region. Moreover, as the system is doped away from half filling, the effective pairing interaction of $d_{xy}$ symmetry is also enhanced and reaches maximum at $\avg{n}\approx0.8$.
Our results also indicate that the system may not exhibit long-range AFM, which is also not observed experimentally \cite{LuJUL9,PhysRevResearch.4.023093,HAYWARD2003839,Cui_2021}.
Although the study of charge correlations does not display a wave vector $q\approx(0.333,0)$, which has been observed in experiments \cite{Tam2022,PhysRevLett.129.027002}, this initial attempt reveals a more complex mechanism should be established to illustrate the CDW phase in nickelates \cite{arXiv:2202.11904}.
In a further work, we simulate the effect of symmetry
breaking by modifying the periodic chemical potential, which shows a different CDW pattern \cite{key1}.
All in all, our work shares exact numerical results to understand the superconducting and symmetry-breaking orders of nickelate-based materials.

%\noindent
%\underline{\it Summary}---

\noindent
\section{Acknowledgements}
We thank Huijia Dai and Jingyao Meng for useful discussions.
This work was supported by NSFC (Grant No. 11974049).
The numerical simulations in this work were performed at the HSCC of
Beijing Normal University and Tianhe in Beijing Computational Science Research Center.

\appendix

\section{Appendix}

In this appendix, we present detailed information on hopping parameters for our Wannier downfolding, the pairing symmetries of the nickel square,
correction of the Trotter error, and the sign problem.

\section{1. Hopping parameters for our Wannier downfolding}
In consideration of the two-band model in our Hamiltonian,
we chose orbital sets of $Ni_{3d_{x^{2} \mbox{-} y^{2}} }$ and $Nd/La_{5d_{z^2}}$ in
Wannier downfolding calculations as implemented in Wannier90 \cite{MOSTOFI2008685}, which can reproduce the
band structure near the Fermi level accurately. The calculated hopping parameters for
two-orbital Wannierization are listed in Table~\ref{table2}.

\begin{table}[htbp]
    \begin{tabular}{p{1cm}<{\centering} p{1cm}<{\centering} p{1cm}<{\centering}
         p{2cm}<{\centering} p{2cm}<{\centering}}
        \hline\hline
        \noalign{\smallskip}
        \multicolumn{5}{c}{Hopping parameters for the tight binding model}\\
        \noalign{\smallskip}
        \hline
        \noalign{\smallskip}
        \multicolumn{3}{c}{} & $Nd Ni O_2$ & $La Ni O_2$ \\
        \noalign{\smallskip}
        \hline
        \noalign{\smallskip}
        i & j & k & \multicolumn{2}{c}{$t^{Ni}_{\left[i, j, k\right]}$} \\
        \noalign{\smallskip}
        \hline
        \noalign{\smallskip}
        0 & 0 & 0 & 0.306385 & 0.284621 \\
        1 & 0 & 0 & -0.377362 & -0.380994 \\
        1 & 1 & 0 & 0.094731 & 0.095830 \\
        2 & 0 & 0 & -0.049510 & -0.049076 \\
        0 & 0 & 1 & -0.027912 & -0.032524 \\
        1 & 0 & 1 & -0.001615 & 0.000423 \\
        1 & 1 & 1 & 0.008920 & 0.009345 \\
        0 & 0 & 2 & 0.001415 & 0.000151 \\
        0 & 0 & 3 & -0.000053 & 0.001201 \\
        \noalign{\smallskip}
        %\hline
        %\noalign{\smallskip}
        i & j & k & \multicolumn{2}{c}{$t^{Nd/La}_{\left[i, j, k\right]}$} \\
        %\noalign{\smallskip}
        %\hline
        \noalign{\smallskip}
        0 & 0 & 0 & 1.493987 & 1.219156 \\
        1 & 0 & 0 & -0.02938 & -0.068788 \\
        1 & 1 & 0 & -0.157513 & -0.087446 \\
        2 & 0 & 0 & 0.051356 & 0.021989 \\
        0 & 0 & 1 & -0.293301 & -0.048961 \\
        1 & 0 & 1 & 0.015698 & -0.196251 \\
        1 & 1 & 1 & 0.004019 & -0.005498 \\
        0 & 0 & 2 & 0.027121 & -0.099677 \\
        0 & 0 & 3 & 0.006169 & -0.003715 \\
        \noalign{\smallskip}
        %\hline
        %\noalign{\smallskip}
        i & j & k & \multicolumn{2}{c}{$t^{Nd/La \mbox{-} Ni}_{\left[i, j, k\right]}$} \\
        %\noalign{\smallskip}
        %\hline
        \noalign{\smallskip}
        0 & 0 & 0 & 0.000151 & -0.011252 \\
        1 & 0 & 0 & -0.000239 & 0.009664 \\
        1 & 1 & 0 & -0.000037 & 0.003106 \\
        2 & 0 & 0 & 0.020577 & 0.001579 \\
        0 & 0 & 1 & -0.000157  & -0.006010  \\
        1 & 0 & 1 & -0.007317 & 0.008403 \\
        1 & 1 & 1 & 0.000070 & 0.006332 \\
        0 & 0 & 2 & 0.000102 & -0.004345 \\
        0 & 0 & 3 & -0.000040 & -0.001681 \\
        \hline\hline
    \end{tabular}
    \caption{On-site energy and hopping parameters (eV) for two-orbital wannierization for $Nd Ni O_2$ and $La Ni O_2$.}
    \label{table2}
\end{table}

\section{2. The pairing symmetries of the nickel-square}
\begin{figure}[tbp]
\includegraphics[scale=0.425, trim = 15 200 0 450, clip]{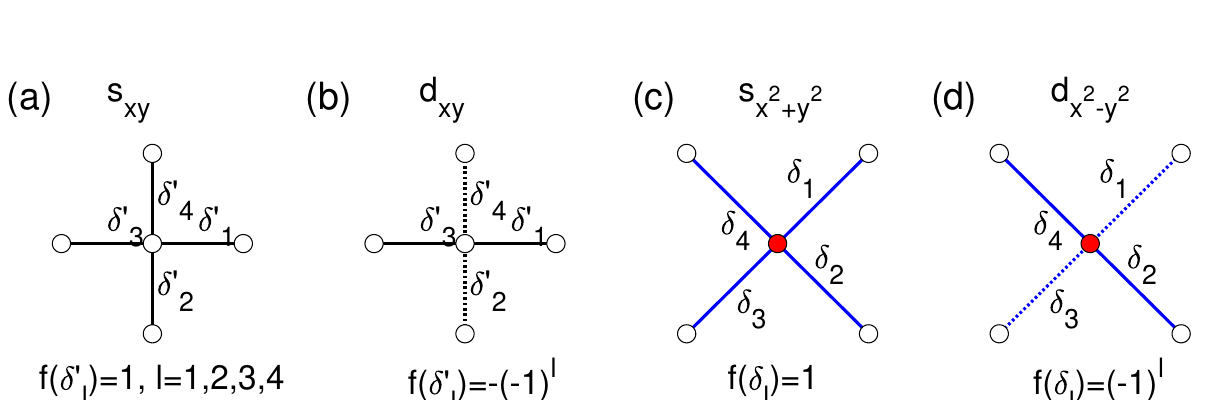}
\caption{(Color online) Phase of the $s_{xy}$, $d_{xy}$, $s_{x^2+y^2}$ and $d_{x^2-y^2}$.
}
\label{Fig:Pairing}
\end{figure}

We referenced four kinds of pairing forms from the iron-square lattice\cite{PhysRevLett.110.107002}, which are pictured in Fig.~\ref{Fig:Pairing}.
These singlet $s$-wave and $d$-wave pairings have the form factor
\begin{eqnarray}
&\text{$s_{xy}$-wave}&: f_{s_{xy}}(\delta'_{l})=1,~l=1,2,3,4,  \notag \\
&\text{$d_{xy}$-wave}&: f_{d_{xy}}(\delta'_{l})=1  ({\delta'_{l}}=(\pm 2\hat{x},0)) \notag \\
&\text{and}& f_{d_{xy}}({\delta'_{l}})=-1  ({\delta'_{l}}=(0,\pm 2\hat{y})),  \notag \\
&\text{$s_{x^2+y^2}$-wave}&: f_{s_{x^2+y^2}}({\delta'_{l}})=1,~l=1,2,3,4,\notag\\
&\text{$d_{x^2-y^2}$-wave}&: f_{d_{x^2-y^2}}({\delta'_{l}})=1  ({\delta'_{l}}=\pm(-\hat{x},\hat{y})) \notag \\
&\text{and}& f_{d_{x^2-y^2}}({\delta'_{l}})=-1  ({\delta'_{l}}=\pm(\hat{x},\hat{y})).
\end{eqnarray}

In experiment, by using scanning tunneling microscopy\cite{science.aal1575} or high-resolution 
laser-ARPES \cite{PhysRevB.79.024517,Ai_2019}, there may be a way to distinguish the $d_{xy}$ and $d_{x^2-y^2}$ pairings.

\begin{figure}[tbp]
\includegraphics[scale=0.29,trim = 200 40 160 10, clip]{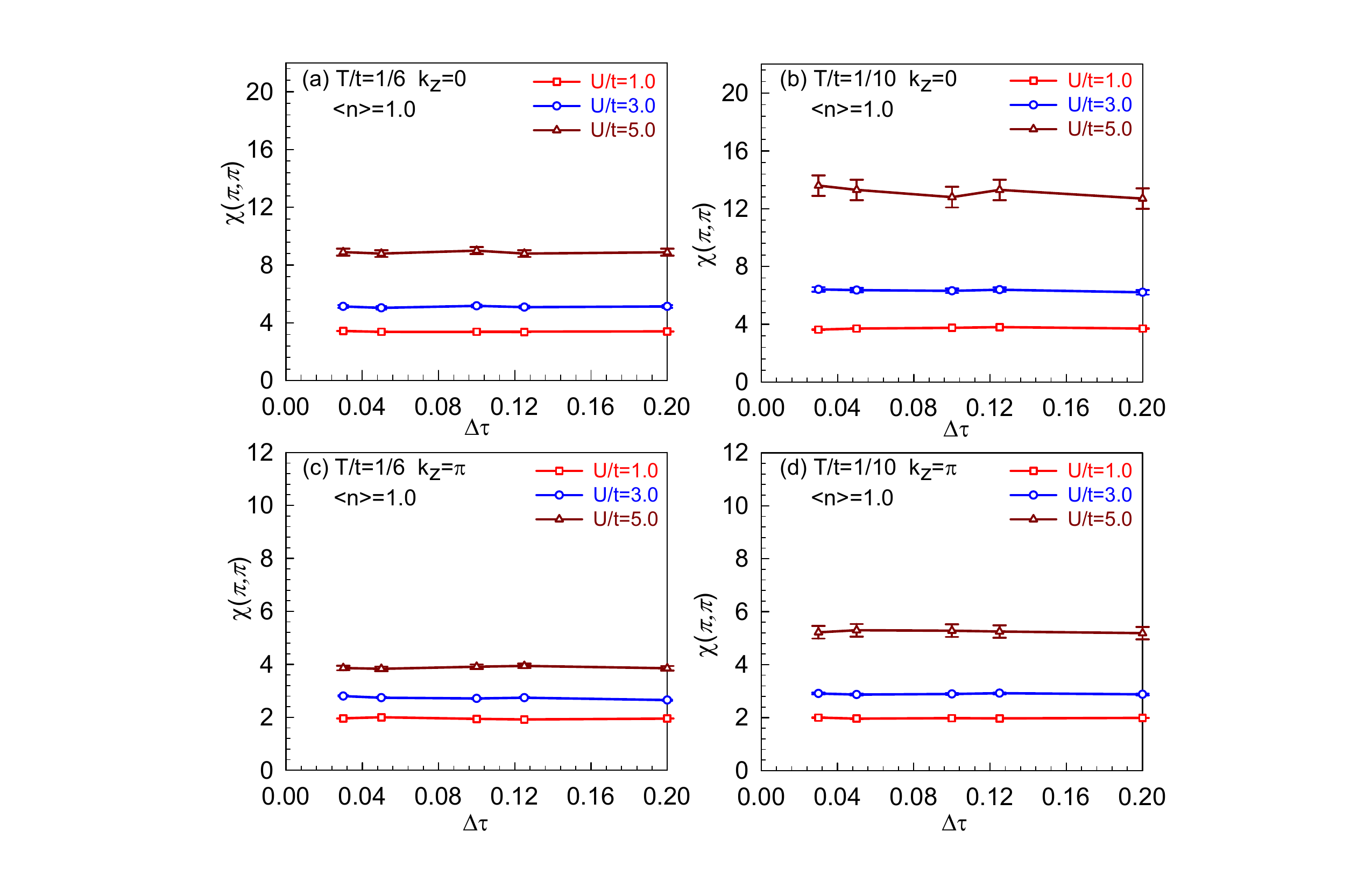}
\caption{(Color online) Influence of the imaginary time step $\Delta\tau$ to the $(\pi,\pi)$ antiferromagnetic correlation $\chi(\pi,\pi)$ for different $U$ at (a) $T/t=1/6$, $k_z=0$, (b) $T/t=1/10$, $k_z=0$, (c) $T/t=1/6$, $k_z=\pi$, (d) $T/t=1/10$, $k_z=\pi$ on a $2\times8^2$ lattice.}
\label{Fig:Figdetau}
\end{figure}

\section{3. Correction of the Trotter error}
Since the operators $H_K$ (kinetic energy) and $H_U$ (potential energy) do not commute, the DQMC algorithm employs the Trotter-Suzuki decomposition to approximate the partition function and then the imaginary-time propagator can be written as
\begin{equation}
  e^{-\Delta\tau H} \approx e^{-\Delta\tau {H_K}}\,e^{-\Delta\tau {H_U}},
\end{equation}

In this process, we can correct systematic error by extrapolating the results at different time steps to the $\Delta\tau=0$ limit. In Fig.~\ref{Fig:Figdetau}, 
we show an impact of the imaginary time step $\Delta\tau$ to the $(\pi,\pi)$ antiferromagnetic correlation $\chi(\pi,\pi)$. The figure indicates that, regardless 
of the interaction strength and temperature, the $\chi(\pi,\pi)$ is essentially identical within the different time steps. Other observables can see a similar 
behavior. As such, Trotter errors can be negligible at the $\Delta\tau$ value used in this paper.

\section{4. The Sign problem}
\begin{figure}[tbp]
\includegraphics[scale=0.4,trim = 60 480 60 0, clip]{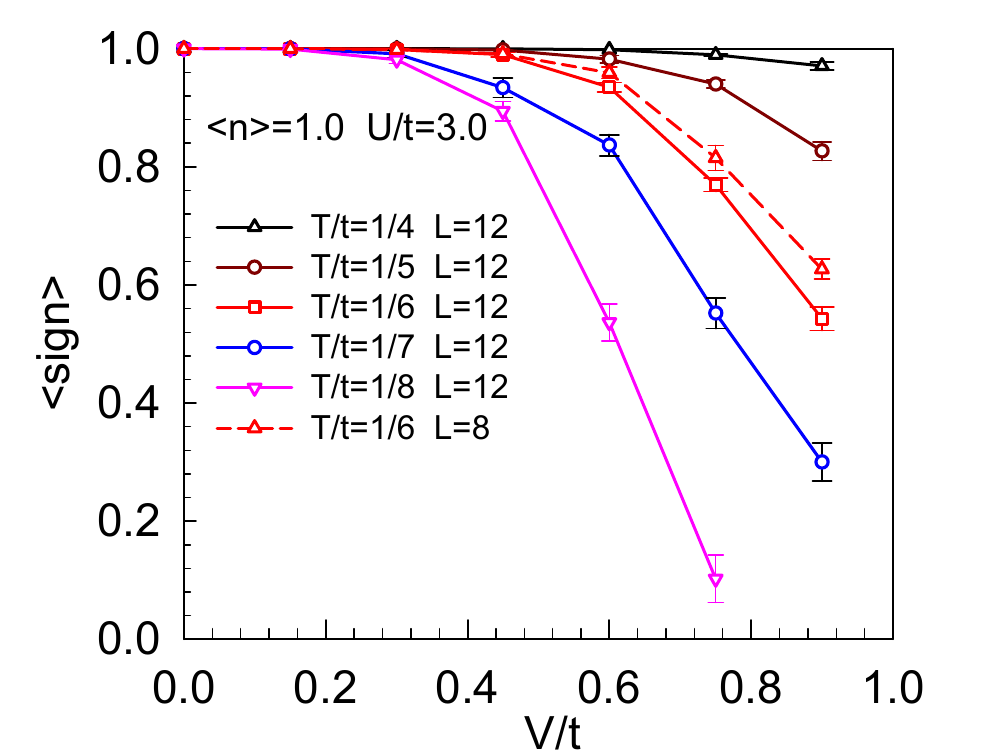}
\caption{(Color online) Average sign $\avg{sign}$ as a function of nearest-neighbor interaction $V$ for different temperatures at $\avg{n}=1.0$, $U/t=3.0$ and $k_z=0$ on $2\times12^2$ or $2\times8^2$ lattice.}
\label{Fig:Figsign}
\end{figure}

For the finite-temperature DQMC method, the infamous sign problem prevents accuracy of results for higher interaction, lower temperature, and larger lattice. Therefore, we assess the average of sign carefully.
In our simulations, the pure on-site interaction does not make the sign-problem terrible for different electron fillings even at low temperatures (and $\avg{sign}\approx 1$).
However, we found that the sign problem became worse when we consider the 
nearest-neighbor repulsion of the Ni 3$d$ orbital to compute the charge-density-wave (CDW) state.
Fig.~\ref{Fig:Figsign} shows the effect of the nearest-neighbor interaction and the temperature on the sign problem 
with measurements of 10000 times.
We can notice that the sign problem becomes worse with increasing 
interaction or decreasing temperature. Our present results are reliable because the average sign is still larger 
than 0.50 for $V=0.9t$, $U/t = 3.0$, and $T/t = 1/6$ on an $L=12$ lattice. To keep the same quality of data with $\avg{sign}\approx 1$,
much longer measurements are essential to compensate the fluctuations. In fact,
the measurements should be enlarged by a factor on the order of $\avg{sign}^{-2}$\cite{PhysRevD.24.2278,SANTOS2003}.
In our simulations, we have made measurement of more than 40000 times for some results.
Therefore, the results with the current Monte Carlo parameters are reliable.
%In order to maintain accuracy, we choose proper $V$ and $T$ where the sign problem is not severe.

\bibliography{reference}

\begin{thebibliography}{73}%
\makeatletter
\providecommand \@ifxundefined [1]{%
 \@ifx{#1\undefined}
}%
\providecommand \@ifnum [1]{%
 \ifnum #1\expandafter \@firstoftwo
 \else \expandafter \@secondoftwo
 \fi
}%
\providecommand \@ifx [1]{%
 \ifx #1\expandafter \@firstoftwo
 \else \expandafter \@secondoftwo
 \fi
}%
\providecommand \natexlab [1]{#1}%
\providecommand \enquote  [1]{``#1''}%
\providecommand \bibnamefont  [1]{#1}%
\providecommand \bibfnamefont [1]{#1}%
\providecommand \citenamefont [1]{#1}%
\providecommand \href@noop [0]{\@secondoftwo}%
\providecommand \href [0]{\begingroup \@sanitize@url \@href}%
\providecommand \@href[1]{\@@startlink{#1}\@@href}%
\providecommand \@@href[1]{\endgroup#1\@@endlink}%
\providecommand \@sanitize@url [0]{\catcode `\\12\catcode `\$12\catcode
  `\&12\catcode `\#12\catcode `\^12\catcode `\_12\catcode `\%12\relax}%
\providecommand \@@startlink[1]{}%
\providecommand \@@endlink[0]{}%
\providecommand \url  [0]{\begingroup\@sanitize@url \@url }%
\providecommand \@url [1]{\endgroup\@href {#1}{\urlprefix }}%
\providecommand \urlprefix  [0]{URL }%
\providecommand \Eprint [0]{\href }%
\providecommand \doibase [0]{http://dx.doi.org/}%
\providecommand \selectlanguage [0]{\@gobble}%
\providecommand \bibinfo  [0]{\@secondoftwo}%
\providecommand \bibfield  [0]{\@secondoftwo}%
\providecommand \translation [1]{[#1]}%
\providecommand \BibitemOpen [0]{}%
\providecommand \bibitemStop [0]{}%
\providecommand \bibitemNoStop [0]{.\EOS\space}%
\providecommand \EOS [0]{\spacefactor3000\relax}%
\providecommand \BibitemShut  [1]{\csname bibitem#1\endcsname}%
\let\auto@bib@innerbib\@empty
%</preamble>
\bibitem [{\citenamefont {Bardeen}\ \emph {et~al.}(1957)\citenamefont
  {Bardeen}, \citenamefont {Cooper},\ and\ \citenamefont
  {Schrieffer}}]{PhysRev.108.1175}%
  \BibitemOpen
  \bibfield  {author} {\bibinfo {author} {\bibfnamefont {J.}~\bibnamefont
  {Bardeen}}, \bibinfo {author} {\bibfnamefont {L.~N.}\ \bibnamefont {Cooper}},
  \ and\ \bibinfo {author} {\bibfnamefont {J.~R.}\ \bibnamefont {Schrieffer}},\
  }\href {\doibase 10.1103/PhysRev.108.1175} {\bibfield  {journal} {\bibinfo
  {journal} {Phys. Rev.}\ }\textbf {\bibinfo {volume} {108}},\ \bibinfo {pages}
  {1175} (\bibinfo {year} {1957})}\BibitemShut {NoStop}%
\bibitem [{\citenamefont {Bednorz}\ and\ \citenamefont
  {M{\"u}ller}(1986)}]{Bednorz1986}%
  \BibitemOpen
  \bibfield  {author} {\bibinfo {author} {\bibfnamefont {J.~G.}\ \bibnamefont
  {Bednorz}}\ and\ \bibinfo {author} {\bibfnamefont {K.~A.}\ \bibnamefont
  {M{\"u}ller}},\ }\href {\doibase 10.1007/BF01303701} {\bibfield  {journal}
  {\bibinfo  {journal} {Zeitschrift f{\"u}r Physik B Condensed Matter}\
  }\textbf {\bibinfo {volume} {64}},\ \bibinfo {pages} {189} (\bibinfo {year}
  {1986})}\BibitemShut {NoStop}%
\bibitem [{\citenamefont {Anderson}(1987)}]{doi:10.1126/science.235.4793.1196}%
  \BibitemOpen
  \bibfield  {author} {\bibinfo {author} {\bibfnamefont {P.~W.}\ \bibnamefont
  {Anderson}},\ }\href {\doibase 10.1126/science.235.4793.1196} {\bibfield
  {journal} {\bibinfo  {journal} {Science}\ }\textbf {\bibinfo {volume}
  {235}},\ \bibinfo {pages} {1196} (\bibinfo {year} {1987})}\BibitemShut
  {NoStop}%
\bibitem [{\citenamefont {Bednorz}\ and\ \citenamefont
  {M\"uller}(1988)}]{RevModPhys.60.585}%
  \BibitemOpen
  \bibfield  {author} {\bibinfo {author} {\bibfnamefont {J.~G.}\ \bibnamefont
  {Bednorz}}\ and\ \bibinfo {author} {\bibfnamefont {K.~A.}\ \bibnamefont
  {M\"uller}},\ }\href {\doibase 10.1103/RevModPhys.60.585} {\bibfield
  {journal} {\bibinfo  {journal} {Rev. Mod. Phys.}\ }\textbf {\bibinfo {volume}
  {60}},\ \bibinfo {pages} {585} (\bibinfo {year} {1988})}\BibitemShut
  {NoStop}%
\bibitem [{\citenamefont {Anderson}\ \emph {et~al.}(2004)\citenamefont
  {Anderson}, \citenamefont {Lee}, \citenamefont {Randeria}, \citenamefont
  {Rice}, \citenamefont {Trivedi},\ and\ \citenamefont
  {Zhang}}]{Anderson_2004}%
  \BibitemOpen
  \bibfield  {author} {\bibinfo {author} {\bibfnamefont {P.~W.}\ \bibnamefont
  {Anderson}}, \bibinfo {author} {\bibfnamefont {P.~A.}\ \bibnamefont {Lee}},
  \bibinfo {author} {\bibfnamefont {M.}~\bibnamefont {Randeria}}, \bibinfo
  {author} {\bibfnamefont {T.~M.}\ \bibnamefont {Rice}}, \bibinfo {author}
  {\bibfnamefont {N.}~\bibnamefont {Trivedi}}, \ and\ \bibinfo {author}
  {\bibfnamefont {F.~C.}\ \bibnamefont {Zhang}},\ }\href {\doibase
  10.1088/0953-8984/16/24/r02} {\bibfield  {journal} {\bibinfo  {journal}
  {Journal of Physics: Condensed Matter}\ }\textbf {\bibinfo {volume} {16}},\
  \bibinfo {pages} {R755} (\bibinfo {year} {2004})}\BibitemShut {NoStop}%
\bibitem [{\citenamefont {Lee}\ \emph {et~al.}(2006)\citenamefont {Lee},
  \citenamefont {Nagaosa},\ and\ \citenamefont {Wen}}]{RevModPhys.78.17}%
  \BibitemOpen
  \bibfield  {author} {\bibinfo {author} {\bibfnamefont {P.~A.}\ \bibnamefont
  {Lee}}, \bibinfo {author} {\bibfnamefont {N.}~\bibnamefont {Nagaosa}}, \ and\
  \bibinfo {author} {\bibfnamefont {X.-G.}\ \bibnamefont {Wen}},\ }\href
  {\doibase 10.1103/RevModPhys.78.17} {\bibfield  {journal} {\bibinfo
  {journal} {Rev. Mod. Phys.}\ }\textbf {\bibinfo {volume} {78}},\ \bibinfo
  {pages} {17} (\bibinfo {year} {2006})}\BibitemShut {NoStop}%
\bibitem [{\citenamefont {Scalapino}(2012)}]{RevModPhys.84.1383}%
  \BibitemOpen
  \bibfield  {author} {\bibinfo {author} {\bibfnamefont {D.~J.}\ \bibnamefont
  {Scalapino}},\ }\href {\doibase 10.1103/RevModPhys.84.1383} {\bibfield
  {journal} {\bibinfo  {journal} {Rev. Mod. Phys.}\ }\textbf {\bibinfo {volume}
  {84}},\ \bibinfo {pages} {1383} (\bibinfo {year} {2012})}\BibitemShut
  {NoStop}%
\bibitem [{\citenamefont {Fradkin}\ \emph {et~al.}(2015)\citenamefont
  {Fradkin}, \citenamefont {Kivelson},\ and\ \citenamefont
  {Tranquada}}]{RevModPhys.87.457}%
  \BibitemOpen
  \bibfield  {author} {\bibinfo {author} {\bibfnamefont {E.}~\bibnamefont
  {Fradkin}}, \bibinfo {author} {\bibfnamefont {S.~A.}\ \bibnamefont
  {Kivelson}}, \ and\ \bibinfo {author} {\bibfnamefont {J.~M.}\ \bibnamefont
  {Tranquada}},\ }\href {\doibase 10.1103/RevModPhys.87.457} {\bibfield
  {journal} {\bibinfo  {journal} {Rev. Mod. Phys.}\ }\textbf {\bibinfo {volume}
  {87}},\ \bibinfo {pages} {457} (\bibinfo {year} {2015})}\BibitemShut
  {NoStop}%
\bibitem [{\citenamefont {Comin}\ and\ \citenamefont
  {Damascelli}(2016)}]{annurev-conmatphys}%
  \BibitemOpen
  \bibfield  {author} {\bibinfo {author} {\bibfnamefont {R.}~\bibnamefont
  {Comin}}\ and\ \bibinfo {author} {\bibfnamefont {A.}~\bibnamefont
  {Damascelli}},\ }\href {\doibase 10.1146/annurev-conmatphys-031115-011401}
  {\bibfield  {journal} {\bibinfo  {journal} {Annual Review of Condensed Matter
  Physics}\ }\textbf {\bibinfo {volume} {7}},\ \bibinfo {pages} {369} (\bibinfo
  {year} {2016})}\BibitemShut {NoStop}%
\bibitem [{\citenamefont {Keimer}\ \emph {et~al.}(2015)\citenamefont {Keimer},
  \citenamefont {Kivelson}, \citenamefont {Norman}, \citenamefont {Uchida},\
  and\ \citenamefont {Zaanen}}]{Keimer2015}%
  \BibitemOpen
  \bibfield  {author} {\bibinfo {author} {\bibfnamefont {B.}~\bibnamefont
  {Keimer}}, \bibinfo {author} {\bibfnamefont {S.~A.}\ \bibnamefont
  {Kivelson}}, \bibinfo {author} {\bibfnamefont {M.~R.}\ \bibnamefont
  {Norman}}, \bibinfo {author} {\bibfnamefont {S.}~\bibnamefont {Uchida}}, \
  and\ \bibinfo {author} {\bibfnamefont {J.}~\bibnamefont {Zaanen}},\ }\href
  {\doibase 10.1038/nature14165} {\bibfield  {journal} {\bibinfo  {journal}
  {Nature}\ }\textbf {\bibinfo {volume} {518}},\ \bibinfo {pages} {179}
  (\bibinfo {year} {2015})}\BibitemShut {NoStop}%
\bibitem [{\citenamefont {Li}\ \emph {et~al.}(2019)\citenamefont {Li},
  \citenamefont {Lee}, \citenamefont {Wang}, \citenamefont {Osada},
  \citenamefont {Crossley}, \citenamefont {Lee}, \citenamefont {Cui},
  \citenamefont {Hikita},\ and\ \citenamefont {Hwang}}]{Li2019}%
  \BibitemOpen
  \bibfield  {author} {\bibinfo {author} {\bibfnamefont {D.}~\bibnamefont
  {Li}}, \bibinfo {author} {\bibfnamefont {K.}~\bibnamefont {Lee}}, \bibinfo
  {author} {\bibfnamefont {B.~Y.}\ \bibnamefont {Wang}}, \bibinfo {author}
  {\bibfnamefont {M.}~\bibnamefont {Osada}}, \bibinfo {author} {\bibfnamefont
  {S.}~\bibnamefont {Crossley}}, \bibinfo {author} {\bibfnamefont {H.~R.}\
  \bibnamefont {Lee}}, \bibinfo {author} {\bibfnamefont {Y.}~\bibnamefont
  {Cui}}, \bibinfo {author} {\bibfnamefont {Y.}~\bibnamefont {Hikita}}, \ and\
  \bibinfo {author} {\bibfnamefont {H.~Y.}\ \bibnamefont {Hwang}},\ }\href
  {\doibase 10.1038/s41586-019-1496-5} {\bibfield  {journal} {\bibinfo
  {journal} {Nature}\ }\textbf {\bibinfo {volume} {572}},\ \bibinfo {pages}
  {624} (\bibinfo {year} {2019})}\BibitemShut {NoStop}%
\bibitem [{\citenamefont {Sawatzky}(2019)}]{SawatzkyAUG29}%
  \BibitemOpen
  \bibfield  {author} {\bibinfo {author} {\bibfnamefont {G.~A.}\ \bibnamefont
  {Sawatzky}},\ }\href {\doibase 10.1038/d41586-019-02518-3} {\bibfield
  {journal} {\bibinfo  {journal} {Nature}\ }\textbf {\bibinfo {volume} {572}},\
  \bibinfo {pages} {592} (\bibinfo {year} {2019})}\BibitemShut {NoStop}%
\bibitem [{\citenamefont {Zeng}\ \emph {et~al.}(2020)\citenamefont {Zeng},
  \citenamefont {Tang}, \citenamefont {Yin}, \citenamefont {Li}, \citenamefont
  {Li}, \citenamefont {Huang}, \citenamefont {Hu}, \citenamefont {Liu},
  \citenamefont {Omar}, \citenamefont {Jani}, \citenamefont {Lim},
  \citenamefont {Han}, \citenamefont {Wan}, \citenamefont {Yang}, \citenamefont
  {Pennycook}, \citenamefont {Wee},\ and\ \citenamefont
  {Ariando}}]{PhysRevLett.125.147003}%
  \BibitemOpen
  \bibfield  {author} {\bibinfo {author} {\bibfnamefont {S.}~\bibnamefont
  {Zeng}}, \bibinfo {author} {\bibfnamefont {C.~S.}\ \bibnamefont {Tang}},
  \bibinfo {author} {\bibfnamefont {X.}~\bibnamefont {Yin}}, \bibinfo {author}
  {\bibfnamefont {C.}~\bibnamefont {Li}}, \bibinfo {author} {\bibfnamefont
  {M.}~\bibnamefont {Li}}, \bibinfo {author} {\bibfnamefont {Z.}~\bibnamefont
  {Huang}}, \bibinfo {author} {\bibfnamefont {J.}~\bibnamefont {Hu}}, \bibinfo
  {author} {\bibfnamefont {W.}~\bibnamefont {Liu}}, \bibinfo {author}
  {\bibfnamefont {G.~J.}\ \bibnamefont {Omar}}, \bibinfo {author}
  {\bibfnamefont {H.}~\bibnamefont {Jani}}, \bibinfo {author} {\bibfnamefont
  {Z.~S.}\ \bibnamefont {Lim}}, \bibinfo {author} {\bibfnamefont
  {K.}~\bibnamefont {Han}}, \bibinfo {author} {\bibfnamefont {D.}~\bibnamefont
  {Wan}}, \bibinfo {author} {\bibfnamefont {P.}~\bibnamefont {Yang}}, \bibinfo
  {author} {\bibfnamefont {S.~J.}\ \bibnamefont {Pennycook}}, \bibinfo {author}
  {\bibfnamefont {A.~T.~S.}\ \bibnamefont {Wee}}, \ and\ \bibinfo {author}
  {\bibfnamefont {A.}~\bibnamefont {Ariando}},\ }\href {\doibase
  10.1103/PhysRevLett.125.147003} {\bibfield  {journal} {\bibinfo  {journal}
  {Phys. Rev. Lett.}\ }\textbf {\bibinfo {volume} {125}},\ \bibinfo {pages}
  {147003} (\bibinfo {year} {2020})}\BibitemShut {NoStop}%
\bibitem [{\citenamefont {Chow}\ \emph {et~al.}(2022)\citenamefont {Chow},
  \citenamefont {Sudheesh}, \citenamefont {Nandi}, \citenamefont {Zeng},
  \citenamefont {Zhang}, \citenamefont {Du}, \citenamefont {Lim}, \citenamefont
  {Chia},\ and\ \citenamefont {Ariando}}]{arXiv:2201.10038}%
  \BibitemOpen
  \bibfield  {author} {\bibinfo {author} {\bibfnamefont {L.~E.}\ \bibnamefont
  {Chow}}, \bibinfo {author} {\bibfnamefont {S.~K.}\ \bibnamefont {Sudheesh}},
  \bibinfo {author} {\bibfnamefont {P.}~\bibnamefont {Nandi}}, \bibinfo
  {author} {\bibfnamefont {S.~W.}\ \bibnamefont {Zeng}}, \bibinfo {author}
  {\bibfnamefont {Z.~T.}\ \bibnamefont {Zhang}}, \bibinfo {author}
  {\bibfnamefont {X.~M.}\ \bibnamefont {Du}}, \bibinfo {author} {\bibfnamefont
  {Z.~S.}\ \bibnamefont {Lim}}, \bibinfo {author} {\bibfnamefont {E.~E.~M.}\
  \bibnamefont {Chia}}, \ and\ \bibinfo {author} {\bibfnamefont
  {A.}~\bibnamefont {Ariando}},\ }\href@noop {} {} (\bibinfo {year} {2022}),\
  \Eprint {http://arxiv.org/abs/2201.10038} {arXiv:2201.10038} \BibitemShut
  {NoStop}%
\bibitem [{\citenamefont {Li}\ \emph {et~al.}(2020)\citenamefont {Li},
  \citenamefont {Wang}, \citenamefont {Lee}, \citenamefont {Harvey},
  \citenamefont {Osada}, \citenamefont {Goodge}, \citenamefont {Kourkoutis},\
  and\ \citenamefont {Hwang}}]{PhysRevLett.125.027001}%
  \BibitemOpen
  \bibfield  {author} {\bibinfo {author} {\bibfnamefont {D.}~\bibnamefont
  {Li}}, \bibinfo {author} {\bibfnamefont {B.~Y.}\ \bibnamefont {Wang}},
  \bibinfo {author} {\bibfnamefont {K.}~\bibnamefont {Lee}}, \bibinfo {author}
  {\bibfnamefont {S.~P.}\ \bibnamefont {Harvey}}, \bibinfo {author}
  {\bibfnamefont {M.}~\bibnamefont {Osada}}, \bibinfo {author} {\bibfnamefont
  {B.~H.}\ \bibnamefont {Goodge}}, \bibinfo {author} {\bibfnamefont {L.~F.}\
  \bibnamefont {Kourkoutis}}, \ and\ \bibinfo {author} {\bibfnamefont {H.~Y.}\
  \bibnamefont {Hwang}},\ }\href {\doibase 10.1103/PhysRevLett.125.027001}
  {\bibfield  {journal} {\bibinfo  {journal} {Phys. Rev. Lett.}\ }\textbf
  {\bibinfo {volume} {125}},\ \bibinfo {pages} {027001} (\bibinfo {year}
  {2020})}\BibitemShut {NoStop}%
\bibitem [{\citenamefont {Choi}\ \emph {et~al.}(2020)\citenamefont {Choi},
  \citenamefont {Lee},\ and\ \citenamefont {Pickett}}]{PhysRevB.101.020503}%
  \BibitemOpen
  \bibfield  {author} {\bibinfo {author} {\bibfnamefont {M.-Y.}\ \bibnamefont
  {Choi}}, \bibinfo {author} {\bibfnamefont {K.-W.}\ \bibnamefont {Lee}}, \
  and\ \bibinfo {author} {\bibfnamefont {W.~E.}\ \bibnamefont {Pickett}},\
  }\href {\doibase 10.1103/PhysRevB.101.020503} {\bibfield  {journal} {\bibinfo
   {journal} {Phys. Rev. B}\ }\textbf {\bibinfo {volume} {101}},\ \bibinfo
  {pages} {020503} (\bibinfo {year} {2020})}\BibitemShut {NoStop}%
\bibitem [{\citenamefont {Si}\ \emph {et~al.}(2020)\citenamefont {Si},
  \citenamefont {Xiao}, \citenamefont {Kaufmann}, \citenamefont {Tomczak},
  \citenamefont {Lu}, \citenamefont {Zhong},\ and\ \citenamefont
  {Held}}]{PhysRevLett.124.166402}%
  \BibitemOpen
  \bibfield  {author} {\bibinfo {author} {\bibfnamefont {L.}~\bibnamefont
  {Si}}, \bibinfo {author} {\bibfnamefont {W.}~\bibnamefont {Xiao}}, \bibinfo
  {author} {\bibfnamefont {J.}~\bibnamefont {Kaufmann}}, \bibinfo {author}
  {\bibfnamefont {J.~M.}\ \bibnamefont {Tomczak}}, \bibinfo {author}
  {\bibfnamefont {Y.}~\bibnamefont {Lu}}, \bibinfo {author} {\bibfnamefont
  {Z.}~\bibnamefont {Zhong}}, \ and\ \bibinfo {author} {\bibfnamefont
  {K.}~\bibnamefont {Held}},\ }\href {\doibase 10.1103/PhysRevLett.124.166402}
  {\bibfield  {journal} {\bibinfo  {journal} {Phys. Rev. Lett.}\ }\textbf
  {\bibinfo {volume} {124}},\ \bibinfo {pages} {166402} (\bibinfo {year}
  {2020})}\BibitemShut {NoStop}%
\bibitem [{\citenamefont {Ryee}\ \emph {et~al.}(2020)\citenamefont {Ryee},
  \citenamefont {Yoon}, \citenamefont {Kim}, \citenamefont {Jeong},\ and\
  \citenamefont {Han}}]{PhysRevB.101.064513}%
  \BibitemOpen
  \bibfield  {author} {\bibinfo {author} {\bibfnamefont {S.}~\bibnamefont
  {Ryee}}, \bibinfo {author} {\bibfnamefont {H.}~\bibnamefont {Yoon}}, \bibinfo
  {author} {\bibfnamefont {T.~J.}\ \bibnamefont {Kim}}, \bibinfo {author}
  {\bibfnamefont {M.~Y.}\ \bibnamefont {Jeong}}, \ and\ \bibinfo {author}
  {\bibfnamefont {M.~J.}\ \bibnamefont {Han}},\ }\href {\doibase
  10.1103/PhysRevB.101.064513} {\bibfield  {journal} {\bibinfo  {journal}
  {Phys. Rev. B}\ }\textbf {\bibinfo {volume} {101}},\ \bibinfo {pages}
  {064513} (\bibinfo {year} {2020})}\BibitemShut {NoStop}%
\bibitem [{\citenamefont {Krishna}\ \emph {et~al.}(2020)\citenamefont
  {Krishna}, \citenamefont {LaBollita}, \citenamefont {Fumega}, \citenamefont
  {Pardo},\ and\ \citenamefont {Botana}}]{PhysRevB.102.224506}%
  \BibitemOpen
  \bibfield  {author} {\bibinfo {author} {\bibfnamefont {J.}~\bibnamefont
  {Krishna}}, \bibinfo {author} {\bibfnamefont {H.}~\bibnamefont {LaBollita}},
  \bibinfo {author} {\bibfnamefont {A.~O.}\ \bibnamefont {Fumega}}, \bibinfo
  {author} {\bibfnamefont {V.}~\bibnamefont {Pardo}}, \ and\ \bibinfo {author}
  {\bibfnamefont {A.~S.}\ \bibnamefont {Botana}},\ }\href {\doibase
  10.1103/PhysRevB.102.224506} {\bibfield  {journal} {\bibinfo  {journal}
  {Phys. Rev. B}\ }\textbf {\bibinfo {volume} {102}},\ \bibinfo {pages}
  {224506} (\bibinfo {year} {2020})}\BibitemShut {NoStop}%
\bibitem [{\citenamefont {Osada}\ \emph {et~al.}(2020)\citenamefont {Osada},
  \citenamefont {Wang}, \citenamefont {Lee}, \citenamefont {Li},\ and\
  \citenamefont {Hwang}}]{PhysRevMaterials.4.121801}%
  \BibitemOpen
  \bibfield  {author} {\bibinfo {author} {\bibfnamefont {M.}~\bibnamefont
  {Osada}}, \bibinfo {author} {\bibfnamefont {B.~Y.}\ \bibnamefont {Wang}},
  \bibinfo {author} {\bibfnamefont {K.}~\bibnamefont {Lee}}, \bibinfo {author}
  {\bibfnamefont {D.}~\bibnamefont {Li}}, \ and\ \bibinfo {author}
  {\bibfnamefont {H.~Y.}\ \bibnamefont {Hwang}},\ }\href {\doibase
  10.1103/PhysRevMaterials.4.121801} {\bibfield  {journal} {\bibinfo  {journal}
  {Phys. Rev. Materials}\ }\textbf {\bibinfo {volume} {4}},\ \bibinfo {pages}
  {121801} (\bibinfo {year} {2020})}\BibitemShut {NoStop}%
\bibitem [{\citenamefont {Jiang}\ \emph {et~al.}(2019)\citenamefont {Jiang},
  \citenamefont {Si}, \citenamefont {Liao},\ and\ \citenamefont
  {Zhong}}]{PhysRevB.100.201106}%
  \BibitemOpen
  \bibfield  {author} {\bibinfo {author} {\bibfnamefont {P.}~\bibnamefont
  {Jiang}}, \bibinfo {author} {\bibfnamefont {L.}~\bibnamefont {Si}}, \bibinfo
  {author} {\bibfnamefont {Z.}~\bibnamefont {Liao}}, \ and\ \bibinfo {author}
  {\bibfnamefont {Z.}~\bibnamefont {Zhong}},\ }\href {\doibase
  10.1103/PhysRevB.100.201106} {\bibfield  {journal} {\bibinfo  {journal}
  {Phys. Rev. B}\ }\textbf {\bibinfo {volume} {100}},\ \bibinfo {pages}
  {201106} (\bibinfo {year} {2019})}\BibitemShut {NoStop}%
\bibitem [{\citenamefont {Wang}\ \emph
  {et~al.}(2020{\natexlab{a}})\citenamefont {Wang}, \citenamefont {Zheng},
  \citenamefont {Krivyakina}, \citenamefont {Chmaissem}, \citenamefont {Lopes},
  \citenamefont {Lynn}, \citenamefont {Gallington}, \citenamefont {Ren},
  \citenamefont {Rosenkranz}, \citenamefont {Mitchell},\ and\ \citenamefont
  {Phelan}}]{PhysRevMaterials.4.084409}%
  \BibitemOpen
  \bibfield  {author} {\bibinfo {author} {\bibfnamefont {B.-X.}\ \bibnamefont
  {Wang}}, \bibinfo {author} {\bibfnamefont {H.}~\bibnamefont {Zheng}},
  \bibinfo {author} {\bibfnamefont {E.}~\bibnamefont {Krivyakina}}, \bibinfo
  {author} {\bibfnamefont {O.}~\bibnamefont {Chmaissem}}, \bibinfo {author}
  {\bibfnamefont {P.~P.}\ \bibnamefont {Lopes}}, \bibinfo {author}
  {\bibfnamefont {J.~W.}\ \bibnamefont {Lynn}}, \bibinfo {author}
  {\bibfnamefont {L.~C.}\ \bibnamefont {Gallington}}, \bibinfo {author}
  {\bibfnamefont {Y.}~\bibnamefont {Ren}}, \bibinfo {author} {\bibfnamefont
  {S.}~\bibnamefont {Rosenkranz}}, \bibinfo {author} {\bibfnamefont {J.~F.}\
  \bibnamefont {Mitchell}}, \ and\ \bibinfo {author} {\bibfnamefont
  {D.}~\bibnamefont {Phelan}},\ }\href {\doibase
  10.1103/PhysRevMaterials.4.084409} {\bibfield  {journal} {\bibinfo  {journal}
  {Phys. Rev. Materials}\ }\textbf {\bibinfo {volume} {4}},\ \bibinfo {pages}
  {084409} (\bibinfo {year} {2020}{\natexlab{a}})}\BibitemShut {NoStop}%
\bibitem [{\citenamefont {Lee}\ \emph {et~al.}(2020)\citenamefont {Lee},
  \citenamefont {Goodge}, \citenamefont {Li}, \citenamefont {Osada},
  \citenamefont {Wang}, \citenamefont {Cui}, \citenamefont {Kourkoutis},\ and\
  \citenamefont {Hwang}}]{LeeAPR1}%
  \BibitemOpen
  \bibfield  {author} {\bibinfo {author} {\bibfnamefont {K.}~\bibnamefont
  {Lee}}, \bibinfo {author} {\bibfnamefont {B.~H.}\ \bibnamefont {Goodge}},
  \bibinfo {author} {\bibfnamefont {D.}~\bibnamefont {Li}}, \bibinfo {author}
  {\bibfnamefont {M.}~\bibnamefont {Osada}}, \bibinfo {author} {\bibfnamefont
  {B.~Y.}\ \bibnamefont {Wang}}, \bibinfo {author} {\bibfnamefont
  {Y.}~\bibnamefont {Cui}}, \bibinfo {author} {\bibfnamefont {L.~F.}\
  \bibnamefont {Kourkoutis}}, \ and\ \bibinfo {author} {\bibfnamefont {H.~Y.}\
  \bibnamefont {Hwang}},\ }\href {\doibase 10.1063/5.0005103} {\bibfield
  {journal} {\bibinfo  {journal} {APL Materials}\ }\textbf {\bibinfo {volume}
  {8}},\ \bibinfo {pages} {041107} (\bibinfo {year} {2020})}\BibitemShut
  {NoStop}%
\bibitem [{\citenamefont {Osada}\ \emph {et~al.}(2021)\citenamefont {Osada},
  \citenamefont {Wang}, \citenamefont {Goodge}, \citenamefont {Harvey},
  \citenamefont {Lee}, \citenamefont {Li}, \citenamefont {Kourkoutis},\ and\
  \citenamefont {Hwang}}]{OsadaNOV}%
  \BibitemOpen
  \bibfield  {author} {\bibinfo {author} {\bibfnamefont {M.}~\bibnamefont
  {Osada}}, \bibinfo {author} {\bibfnamefont {B.~Y.}\ \bibnamefont {Wang}},
  \bibinfo {author} {\bibfnamefont {B.~H.}\ \bibnamefont {Goodge}}, \bibinfo
  {author} {\bibfnamefont {S.~P.}\ \bibnamefont {Harvey}}, \bibinfo {author}
  {\bibfnamefont {K.~H.}\ \bibnamefont {Lee}}, \bibinfo {author} {\bibfnamefont
  {D.~F.}\ \bibnamefont {Li}}, \bibinfo {author} {\bibfnamefont {L.~F.}\
  \bibnamefont {Kourkoutis}}, \ and\ \bibinfo {author} {\bibfnamefont {H.~Y.}\
  \bibnamefont {Hwang}},\ }\href {\doibase 10.1002/adma.202104083} {\bibfield
  {journal} {\bibinfo  {journal} {Adv.Mater.}\ }\textbf {\bibinfo {volume}
  {33}},\ \bibinfo {pages} {2104083} (\bibinfo {year} {2021})}\BibitemShut
  {NoStop}%
\bibitem [{\citenamefont {Zeng}\ \emph {et~al.}(2022)\citenamefont {Zeng},
  \citenamefont {Li}, \citenamefont {Chow}, \citenamefont {Cao}, \citenamefont
  {Zhang}, \citenamefont {Tang}, \citenamefont {Yin}, \citenamefont {Lim},
  \citenamefont {Hu}, \citenamefont {Yang},\ and\ \citenamefont
  {Ariando}}]{zeng2021superconductivity}%
  \BibitemOpen
  \bibfield  {author} {\bibinfo {author} {\bibfnamefont {S.}~\bibnamefont
  {Zeng}}, \bibinfo {author} {\bibfnamefont {C.}~\bibnamefont {Li}}, \bibinfo
  {author} {\bibfnamefont {L.~E.}\ \bibnamefont {Chow}}, \bibinfo {author}
  {\bibfnamefont {Y.}~\bibnamefont {Cao}}, \bibinfo {author} {\bibfnamefont
  {Z.}~\bibnamefont {Zhang}}, \bibinfo {author} {\bibfnamefont {C.~S.}\
  \bibnamefont {Tang}}, \bibinfo {author} {\bibfnamefont {X.}~\bibnamefont
  {Yin}}, \bibinfo {author} {\bibfnamefont {Z.~S.}\ \bibnamefont {Lim}},
  \bibinfo {author} {\bibfnamefont {J.}~\bibnamefont {Hu}}, \bibinfo {author}
  {\bibfnamefont {P.}~\bibnamefont {Yang}}, \ and\ \bibinfo {author}
  {\bibfnamefont {A.}~\bibnamefont {Ariando}},\ }\href {\doibase
  10.1126/sciadv.abl9927} {\bibfield  {journal} {\bibinfo  {journal} {Science
  Advances}\ }\textbf {\bibinfo {volume} {8}},\ \bibinfo {pages} {eabl9927}
  (\bibinfo {year} {2022})}\BibitemShut {NoStop}%
\bibitem [{\citenamefont {Gu}\ \emph {et~al.}(2020)\citenamefont {Gu},
  \citenamefont {Li}, \citenamefont {Wan}, \citenamefont {Li}, \citenamefont
  {Guo}, \citenamefont {Yang}, \citenamefont {Li}, \citenamefont {Zhu},
  \citenamefont {Pan}, \citenamefont {Nie},\ and\ \citenamefont
  {Wen}}]{Gu2020}%
  \BibitemOpen
  \bibfield  {author} {\bibinfo {author} {\bibfnamefont {Q.}~\bibnamefont
  {Gu}}, \bibinfo {author} {\bibfnamefont {Y.}~\bibnamefont {Li}}, \bibinfo
  {author} {\bibfnamefont {S.}~\bibnamefont {Wan}}, \bibinfo {author}
  {\bibfnamefont {H.}~\bibnamefont {Li}}, \bibinfo {author} {\bibfnamefont
  {W.}~\bibnamefont {Guo}}, \bibinfo {author} {\bibfnamefont {H.}~\bibnamefont
  {Yang}}, \bibinfo {author} {\bibfnamefont {Q.}~\bibnamefont {Li}}, \bibinfo
  {author} {\bibfnamefont {X.}~\bibnamefont {Zhu}}, \bibinfo {author}
  {\bibfnamefont {X.}~\bibnamefont {Pan}}, \bibinfo {author} {\bibfnamefont
  {Y.}~\bibnamefont {Nie}}, \ and\ \bibinfo {author} {\bibfnamefont {H.-H.}\
  \bibnamefont {Wen}},\ }\href {\doibase 10.1038/s41467-020-19908-1} {\bibfield
   {journal} {\bibinfo  {journal} {Nature Communications}\ }\textbf {\bibinfo
  {volume} {11}},\ \bibinfo {pages} {6027} (\bibinfo {year}
  {2020})}\BibitemShut {NoStop}%
\bibitem [{\citenamefont {Zhang}\ \emph {et~al.}(2021)\citenamefont {Zhang},
  \citenamefont {Zhang}, \citenamefont {Guo},\ and\ \citenamefont
  {Yang}}]{Zhang_2021}%
  \BibitemOpen
  \bibfield  {author} {\bibinfo {author} {\bibfnamefont {M.}~\bibnamefont
  {Zhang}}, \bibinfo {author} {\bibfnamefont {Y.}~\bibnamefont {Zhang}},
  \bibinfo {author} {\bibfnamefont {H.}~\bibnamefont {Guo}}, \ and\ \bibinfo
  {author} {\bibfnamefont {F.}~\bibnamefont {Yang}},\ }\href {\doibase
  10.1088/1674-1056/ac0bb1} {\bibfield  {journal} {\bibinfo  {journal} {Chinese
  Physics B}\ }\textbf {\bibinfo {volume} {30}},\ \bibinfo {pages} {108204}
  (\bibinfo {year} {2021})}\BibitemShut {NoStop}%
\bibitem [{\citenamefont {Zhang}\ and\ \citenamefont
  {Vishwanath}(2020)}]{PhysRevResearch.2.023112}%
  \BibitemOpen
  \bibfield  {author} {\bibinfo {author} {\bibfnamefont {Y.-H.}\ \bibnamefont
  {Zhang}}\ and\ \bibinfo {author} {\bibfnamefont {A.}~\bibnamefont
  {Vishwanath}},\ }\href {\doibase 10.1103/PhysRevResearch.2.023112} {\bibfield
   {journal} {\bibinfo  {journal} {Phys. Rev. Research}\ }\textbf {\bibinfo
  {volume} {2}},\ \bibinfo {pages} {023112} (\bibinfo {year}
  {2020})}\BibitemShut {NoStop}%
\bibitem [{\citenamefont {Kitatani}\ \emph {et~al.}(2020)\citenamefont
  {Kitatani}, \citenamefont {Si}, \citenamefont {Janson}, \citenamefont
  {Arita}, \citenamefont {Zhong},\ and\ \citenamefont {Held}}]{Kitatani2020}%
  \BibitemOpen
  \bibfield  {author} {\bibinfo {author} {\bibfnamefont {M.}~\bibnamefont
  {Kitatani}}, \bibinfo {author} {\bibfnamefont {L.}~\bibnamefont {Si}},
  \bibinfo {author} {\bibfnamefont {O.}~\bibnamefont {Janson}}, \bibinfo
  {author} {\bibfnamefont {R.}~\bibnamefont {Arita}}, \bibinfo {author}
  {\bibfnamefont {Z.}~\bibnamefont {Zhong}}, \ and\ \bibinfo {author}
  {\bibfnamefont {K.}~\bibnamefont {Held}},\ }\href {\doibase
  10.1038/s41535-020-00260-y} {\bibfield  {journal} {\bibinfo  {journal} {npj
  Quantum Materials}\ }\textbf {\bibinfo {volume} {5}},\ \bibinfo {pages} {59}
  (\bibinfo {year} {2020})}\BibitemShut {NoStop}%
\bibitem [{\citenamefont {Adhikary}\ \emph {et~al.}(2020)\citenamefont
  {Adhikary}, \citenamefont {Bandyopadhyay}, \citenamefont {Das}, \citenamefont
  {Dasgupta},\ and\ \citenamefont {Saha-Dasgupta}}]{PhysRevB.102.100501}%
  \BibitemOpen
  \bibfield  {author} {\bibinfo {author} {\bibfnamefont {P.}~\bibnamefont
  {Adhikary}}, \bibinfo {author} {\bibfnamefont {S.}~\bibnamefont
  {Bandyopadhyay}}, \bibinfo {author} {\bibfnamefont {T.}~\bibnamefont {Das}},
  \bibinfo {author} {\bibfnamefont {I.}~\bibnamefont {Dasgupta}}, \ and\
  \bibinfo {author} {\bibfnamefont {T.}~\bibnamefont {Saha-Dasgupta}},\ }\href
  {\doibase 10.1103/PhysRevB.102.100501} {\bibfield  {journal} {\bibinfo
  {journal} {Phys. Rev. B}\ }\textbf {\bibinfo {volume} {102}},\ \bibinfo
  {pages} {100501} (\bibinfo {year} {2020})}\BibitemShut {NoStop}%
\bibitem [{\citenamefont {Wu}\ \emph {et~al.}(2020)\citenamefont {Wu},
  \citenamefont {Di~Sante}, \citenamefont {Schwemmer}, \citenamefont {Hanke},
  \citenamefont {Hwang}, \citenamefont {Raghu},\ and\ \citenamefont
  {Thomale}}]{PhysRevB.101.060504}%
  \BibitemOpen
  \bibfield  {author} {\bibinfo {author} {\bibfnamefont {X.}~\bibnamefont
  {Wu}}, \bibinfo {author} {\bibfnamefont {D.}~\bibnamefont {Di~Sante}},
  \bibinfo {author} {\bibfnamefont {T.}~\bibnamefont {Schwemmer}}, \bibinfo
  {author} {\bibfnamefont {W.}~\bibnamefont {Hanke}}, \bibinfo {author}
  {\bibfnamefont {H.~Y.}\ \bibnamefont {Hwang}}, \bibinfo {author}
  {\bibfnamefont {S.}~\bibnamefont {Raghu}}, \ and\ \bibinfo {author}
  {\bibfnamefont {R.}~\bibnamefont {Thomale}},\ }\href {\doibase
  10.1103/PhysRevB.101.060504} {\bibfield  {journal} {\bibinfo  {journal}
  {Phys. Rev. B}\ }\textbf {\bibinfo {volume} {101}},\ \bibinfo {pages}
  {060504} (\bibinfo {year} {2020})}\BibitemShut {NoStop}%
\bibitem [{\citenamefont {Zhang}\ \emph {et~al.}(2020)\citenamefont {Zhang},
  \citenamefont {Yang},\ and\ \citenamefont {Zhang}}]{PhysRevB.101.020501}%
  \BibitemOpen
  \bibfield  {author} {\bibinfo {author} {\bibfnamefont {G.-M.}\ \bibnamefont
  {Zhang}}, \bibinfo {author} {\bibfnamefont {Y.-f.}\ \bibnamefont {Yang}}, \
  and\ \bibinfo {author} {\bibfnamefont {F.-C.}\ \bibnamefont {Zhang}},\ }\href
  {\doibase 10.1103/PhysRevB.101.020501} {\bibfield  {journal} {\bibinfo
  {journal} {Phys. Rev. B}\ }\textbf {\bibinfo {volume} {101}},\ \bibinfo
  {pages} {020501} (\bibinfo {year} {2020})}\BibitemShut {NoStop}%
\bibitem [{\citenamefont {Lu}\ \emph {et~al.}(2022)\citenamefont {Lu},
  \citenamefont {Hu}, \citenamefont {Wang}, \citenamefont {Yang},\ and\
  \citenamefont {Wu}}]{lu2021twoorbital}%
  \BibitemOpen
  \bibfield  {author} {\bibinfo {author} {\bibfnamefont {C.}~\bibnamefont
  {Lu}}, \bibinfo {author} {\bibfnamefont {L.-H.}\ \bibnamefont {Hu}}, \bibinfo
  {author} {\bibfnamefont {Y.}~\bibnamefont {Wang}}, \bibinfo {author}
  {\bibfnamefont {F.}~\bibnamefont {Yang}}, \ and\ \bibinfo {author}
  {\bibfnamefont {C.}~\bibnamefont {Wu}},\ }\href {\doibase
  10.1103/PhysRevB.105.054516} {\bibfield  {journal} {\bibinfo  {journal}
  {Phys. Rev. B}\ }\textbf {\bibinfo {volume} {105}},\ \bibinfo {pages}
  {054516} (\bibinfo {year} {2022})}\BibitemShut {NoStop}%
\bibitem [{\citenamefont {Sakakibara}\ \emph {et~al.}(2020)\citenamefont
  {Sakakibara}, \citenamefont {Usui}, \citenamefont {Suzuki}, \citenamefont
  {Kotani}, \citenamefont {Aoki},\ and\ \citenamefont
  {Kuroki}}]{PhysRevLett.125.077003}%
  \BibitemOpen
  \bibfield  {author} {\bibinfo {author} {\bibfnamefont {H.}~\bibnamefont
  {Sakakibara}}, \bibinfo {author} {\bibfnamefont {H.}~\bibnamefont {Usui}},
  \bibinfo {author} {\bibfnamefont {K.}~\bibnamefont {Suzuki}}, \bibinfo
  {author} {\bibfnamefont {T.}~\bibnamefont {Kotani}}, \bibinfo {author}
  {\bibfnamefont {H.}~\bibnamefont {Aoki}}, \ and\ \bibinfo {author}
  {\bibfnamefont {K.}~\bibnamefont {Kuroki}},\ }\href {\doibase
  10.1103/PhysRevLett.125.077003} {\bibfield  {journal} {\bibinfo  {journal}
  {Phys. Rev. Lett.}\ }\textbf {\bibinfo {volume} {125}},\ \bibinfo {pages}
  {077003} (\bibinfo {year} {2020})}\BibitemShut {NoStop}%
\bibitem [{\citenamefont {Kreisel}\ \emph {et~al.}(2022)\citenamefont
  {Kreisel}, \citenamefont {Andersen}, \citenamefont {R\o{}mer}, \citenamefont
  {Eremin},\ and\ \citenamefont {Lechermann}}]{PhysRevLett.129.077002}%
  \BibitemOpen
  \bibfield  {author} {\bibinfo {author} {\bibfnamefont {A.}~\bibnamefont
  {Kreisel}}, \bibinfo {author} {\bibfnamefont {B.~M.}\ \bibnamefont
  {Andersen}}, \bibinfo {author} {\bibfnamefont {A.~T.}\ \bibnamefont
  {R\o{}mer}}, \bibinfo {author} {\bibfnamefont {I.~M.}\ \bibnamefont
  {Eremin}}, \ and\ \bibinfo {author} {\bibfnamefont {F.}~\bibnamefont
  {Lechermann}},\ }\href {\doibase 10.1103/PhysRevLett.129.077002} {\bibfield
  {journal} {\bibinfo  {journal} {Phys. Rev. Lett.}\ }\textbf {\bibinfo
  {volume} {129}},\ \bibinfo {pages} {077002} (\bibinfo {year}
  {2022})}\BibitemShut {NoStop}%
\bibitem [{\citenamefont {Wang}\ \emph
  {et~al.}(2020{\natexlab{b}})\citenamefont {Wang}, \citenamefont {Zhang},
  \citenamefont {Yang},\ and\ \citenamefont {Zhang}}]{PhysRevB.102.220501}%
  \BibitemOpen
  \bibfield  {author} {\bibinfo {author} {\bibfnamefont {Z.}~\bibnamefont
  {Wang}}, \bibinfo {author} {\bibfnamefont {G.-M.}\ \bibnamefont {Zhang}},
  \bibinfo {author} {\bibfnamefont {Y.-f.}\ \bibnamefont {Yang}}, \ and\
  \bibinfo {author} {\bibfnamefont {F.-C.}\ \bibnamefont {Zhang}},\ }\href
  {\doibase 10.1103/PhysRevB.102.220501} {\bibfield  {journal} {\bibinfo
  {journal} {Phys. Rev. B}\ }\textbf {\bibinfo {volume} {102}},\ \bibinfo
  {pages} {220501} (\bibinfo {year} {2020}{\natexlab{b}})}\BibitemShut
  {NoStop}%
\bibitem [{\citenamefont {Ortiz}\ \emph {et~al.}(2022)\citenamefont {Ortiz},
  \citenamefont {Puphal}, \citenamefont {Klett}, \citenamefont {Hotz},
  \citenamefont {Kremer}, \citenamefont {Trepka}, \citenamefont {Hemmida},
  \citenamefont {von Nidda}, \citenamefont {Isobe}, \citenamefont {Khasanov},
  \citenamefont {Luetkens}, \citenamefont {Hansmann}, \citenamefont {Keimer},
  \citenamefont {Sch\"afer},\ and\ \citenamefont
  {Hepting}}]{PhysRevResearch.4.023093}%
  \BibitemOpen
  \bibfield  {author} {\bibinfo {author} {\bibfnamefont {R.~A.}\ \bibnamefont
  {Ortiz}}, \bibinfo {author} {\bibfnamefont {P.}~\bibnamefont {Puphal}},
  \bibinfo {author} {\bibfnamefont {M.}~\bibnamefont {Klett}}, \bibinfo
  {author} {\bibfnamefont {F.}~\bibnamefont {Hotz}}, \bibinfo {author}
  {\bibfnamefont {R.~K.}\ \bibnamefont {Kremer}}, \bibinfo {author}
  {\bibfnamefont {H.}~\bibnamefont {Trepka}}, \bibinfo {author} {\bibfnamefont
  {M.}~\bibnamefont {Hemmida}}, \bibinfo {author} {\bibfnamefont {H.-A.~K.}\
  \bibnamefont {von Nidda}}, \bibinfo {author} {\bibfnamefont {M.}~\bibnamefont
  {Isobe}}, \bibinfo {author} {\bibfnamefont {R.}~\bibnamefont {Khasanov}},
  \bibinfo {author} {\bibfnamefont {H.}~\bibnamefont {Luetkens}}, \bibinfo
  {author} {\bibfnamefont {P.}~\bibnamefont {Hansmann}}, \bibinfo {author}
  {\bibfnamefont {B.}~\bibnamefont {Keimer}}, \bibinfo {author} {\bibfnamefont
  {T.}~\bibnamefont {Sch\"afer}}, \ and\ \bibinfo {author} {\bibfnamefont
  {M.}~\bibnamefont {Hepting}},\ }\href {\doibase
  10.1103/PhysRevResearch.4.023093} {\bibfield  {journal} {\bibinfo  {journal}
  {Phys. Rev. Research}\ }\textbf {\bibinfo {volume} {4}},\ \bibinfo {pages}
  {023093} (\bibinfo {year} {2022})}\BibitemShut {NoStop}%
\bibitem [{\citenamefont {Lu}\ \emph {et~al.}(2021)\citenamefont {Lu},
  \citenamefont {Rossi}, \citenamefont {Nag}, \citenamefont {Osada},
  \citenamefont {Li}, \citenamefont {Lee}, \citenamefont {Wang}, \citenamefont
  {Garcia-Fernandez}, \citenamefont {Agrestini}, \citenamefont {Shen},
  \citenamefont {Been}, \citenamefont {Moritz}, \citenamefont {Devereaux},
  \citenamefont {Zaanen}, \citenamefont {Hwang}, \citenamefont {Zhou},\ and\
  \citenamefont {Lee}}]{LuJUL9}%
  \BibitemOpen
  \bibfield  {author} {\bibinfo {author} {\bibfnamefont {H.}~\bibnamefont
  {Lu}}, \bibinfo {author} {\bibfnamefont {M.}~\bibnamefont {Rossi}}, \bibinfo
  {author} {\bibfnamefont {A.}~\bibnamefont {Nag}}, \bibinfo {author}
  {\bibfnamefont {M.}~\bibnamefont {Osada}}, \bibinfo {author} {\bibfnamefont
  {D.~F.}\ \bibnamefont {Li}}, \bibinfo {author} {\bibfnamefont
  {K.}~\bibnamefont {Lee}}, \bibinfo {author} {\bibfnamefont {B.~Y.}\
  \bibnamefont {Wang}}, \bibinfo {author} {\bibfnamefont {M.}~\bibnamefont
  {Garcia-Fernandez}}, \bibinfo {author} {\bibfnamefont {S.}~\bibnamefont
  {Agrestini}}, \bibinfo {author} {\bibfnamefont {Z.~X.}\ \bibnamefont {Shen}},
  \bibinfo {author} {\bibfnamefont {E.~M.}\ \bibnamefont {Been}}, \bibinfo
  {author} {\bibfnamefont {B.}~\bibnamefont {Moritz}}, \bibinfo {author}
  {\bibfnamefont {T.~P.}\ \bibnamefont {Devereaux}}, \bibinfo {author}
  {\bibfnamefont {J.}~\bibnamefont {Zaanen}}, \bibinfo {author} {\bibfnamefont
  {H.~Y.}\ \bibnamefont {Hwang}}, \bibinfo {author} {\bibfnamefont {K.-J.}\
  \bibnamefont {Zhou}}, \ and\ \bibinfo {author} {\bibfnamefont {W.~S.}\
  \bibnamefont {Lee}},\ }\href {\doibase 10.1126/science.abd7726} {\bibfield
  {journal} {\bibinfo  {journal} {Science}\ }\textbf {\bibinfo {volume}
  {373}},\ \bibinfo {pages} {213} (\bibinfo {year} {2021})}\BibitemShut
  {NoStop}%
\bibitem [{\citenamefont {Cui}\ \emph {et~al.}(2021)\citenamefont {Cui},
  \citenamefont {Li}, \citenamefont {Li}, \citenamefont {Zhu}, \citenamefont
  {Hu}, \citenamefont {feng Yang}, \citenamefont {Zhang}, \citenamefont {Yu},
  \citenamefont {Wen},\ and\ \citenamefont {Yu}}]{Cui_2021}%
  \BibitemOpen
  \bibfield  {author} {\bibinfo {author} {\bibfnamefont {Y.}~\bibnamefont
  {Cui}}, \bibinfo {author} {\bibfnamefont {C.}~\bibnamefont {Li}}, \bibinfo
  {author} {\bibfnamefont {Q.}~\bibnamefont {Li}}, \bibinfo {author}
  {\bibfnamefont {X.}~\bibnamefont {Zhu}}, \bibinfo {author} {\bibfnamefont
  {Z.}~\bibnamefont {Hu}}, \bibinfo {author} {\bibfnamefont {Y.}~\bibnamefont
  {feng Yang}}, \bibinfo {author} {\bibfnamefont {J.}~\bibnamefont {Zhang}},
  \bibinfo {author} {\bibfnamefont {R.}~\bibnamefont {Yu}}, \bibinfo {author}
  {\bibfnamefont {H.-H.}\ \bibnamefont {Wen}}, \ and\ \bibinfo {author}
  {\bibfnamefont {W.}~\bibnamefont {Yu}},\ }\href {\doibase
  10.1088/0256-307x/38/6/067401} {\bibfield  {journal} {\bibinfo  {journal}
  {Chinese Physics Letters}\ }\textbf {\bibinfo {volume} {38}},\ \bibinfo
  {pages} {067401} (\bibinfo {year} {2021})}\BibitemShut {NoStop}%
\bibitem [{\citenamefont {Rossi}\ \emph {et~al.}(2022)\citenamefont {Rossi},
  \citenamefont {Osada}, \citenamefont {Choi}, \citenamefont {Agrestini},
  \citenamefont {Jost}, \citenamefont {Lee}, \citenamefont {Lu}, \citenamefont
  {Wang}, \citenamefont {Lee}, \citenamefont {Nag}, \citenamefont {Chuang},
  \citenamefont {Kuo}, \citenamefont {Lee}, \citenamefont {Moritz},
  \citenamefont {Devereaux}, \citenamefont {Shen}, \citenamefont {Lee},
  \citenamefont {Zhou}, \citenamefont {Hwang},\ and\ \citenamefont
  {Lee}}]{Rossi2022}%
  \BibitemOpen
  \bibfield  {author} {\bibinfo {author} {\bibfnamefont {M.}~\bibnamefont
  {Rossi}}, \bibinfo {author} {\bibfnamefont {M.}~\bibnamefont {Osada}},
  \bibinfo {author} {\bibfnamefont {J.}~\bibnamefont {Choi}}, \bibinfo {author}
  {\bibfnamefont {S.}~\bibnamefont {Agrestini}}, \bibinfo {author}
  {\bibfnamefont {D.}~\bibnamefont {Jost}}, \bibinfo {author} {\bibfnamefont
  {Y.}~\bibnamefont {Lee}}, \bibinfo {author} {\bibfnamefont {H.}~\bibnamefont
  {Lu}}, \bibinfo {author} {\bibfnamefont {B.~Y.}\ \bibnamefont {Wang}},
  \bibinfo {author} {\bibfnamefont {K.}~\bibnamefont {Lee}}, \bibinfo {author}
  {\bibfnamefont {A.}~\bibnamefont {Nag}}, \bibinfo {author} {\bibfnamefont
  {Y.-D.}\ \bibnamefont {Chuang}}, \bibinfo {author} {\bibfnamefont {C.-T.}\
  \bibnamefont {Kuo}}, \bibinfo {author} {\bibfnamefont {S.-J.}\ \bibnamefont
  {Lee}}, \bibinfo {author} {\bibfnamefont {B.}~\bibnamefont {Moritz}},
  \bibinfo {author} {\bibfnamefont {T.~P.}\ \bibnamefont {Devereaux}}, \bibinfo
  {author} {\bibfnamefont {Z.-X.}\ \bibnamefont {Shen}}, \bibinfo {author}
  {\bibfnamefont {J.-S.}\ \bibnamefont {Lee}}, \bibinfo {author} {\bibfnamefont
  {K.-J.}\ \bibnamefont {Zhou}}, \bibinfo {author} {\bibfnamefont {H.~Y.}\
  \bibnamefont {Hwang}}, \ and\ \bibinfo {author} {\bibfnamefont {W.-S.}\
  \bibnamefont {Lee}},\ }\href {\doibase 10.1038/s41567-022-01660-6} {\bibfield
   {journal} {\bibinfo  {journal} {Nature Physics}\ }\textbf {\bibinfo {volume}
  {18}},\ \bibinfo {pages} {869} (\bibinfo {year} {2022})}\BibitemShut
  {NoStop}%
\bibitem [{\citenamefont {Tam}\ \emph {et~al.}(2022)\citenamefont {Tam},
  \citenamefont {Choi}, \citenamefont {Ding}, \citenamefont {Agrestini},
  \citenamefont {Nag}, \citenamefont {Wu}, \citenamefont {Huang}, \citenamefont
  {Luo}, \citenamefont {Gao}, \citenamefont {Garc{\'i}a-Fern{\'a}ndez},
  \citenamefont {Qiao},\ and\ \citenamefont {Zhou}}]{Tam2022}%
  \BibitemOpen
  \bibfield  {author} {\bibinfo {author} {\bibfnamefont {C.~C.}\ \bibnamefont
  {Tam}}, \bibinfo {author} {\bibfnamefont {J.}~\bibnamefont {Choi}}, \bibinfo
  {author} {\bibfnamefont {X.}~\bibnamefont {Ding}}, \bibinfo {author}
  {\bibfnamefont {S.}~\bibnamefont {Agrestini}}, \bibinfo {author}
  {\bibfnamefont {A.}~\bibnamefont {Nag}}, \bibinfo {author} {\bibfnamefont
  {M.}~\bibnamefont {Wu}}, \bibinfo {author} {\bibfnamefont {B.}~\bibnamefont
  {Huang}}, \bibinfo {author} {\bibfnamefont {H.}~\bibnamefont {Luo}}, \bibinfo
  {author} {\bibfnamefont {P.}~\bibnamefont {Gao}}, \bibinfo {author}
  {\bibfnamefont {M.}~\bibnamefont {Garc{\'i}a-Fern{\'a}ndez}}, \bibinfo
  {author} {\bibfnamefont {L.}~\bibnamefont {Qiao}}, \ and\ \bibinfo {author}
  {\bibfnamefont {K.-J.}\ \bibnamefont {Zhou}},\ }\href {\doibase
  10.1038/s41563-022-01330-1} {\bibfield  {journal} {\bibinfo  {journal}
  {Nature Materials}\ }\textbf {\bibinfo {volume} {21}},\ \bibinfo {pages}
  {1116} (\bibinfo {year} {2022})}\BibitemShut {NoStop}%
\bibitem [{\citenamefont {Krieger}\ \emph {et~al.}(2022)\citenamefont
  {Krieger}, \citenamefont {Martinelli}, \citenamefont {Zeng}, \citenamefont
  {Chow}, \citenamefont {Kummer}, \citenamefont {Arpaia}, \citenamefont
  {Moretti~Sala}, \citenamefont {Brookes}, \citenamefont {Ariando},
  \citenamefont {Viart}, \citenamefont {Salluzzo}, \citenamefont
  {Ghiringhelli},\ and\ \citenamefont {Preziosi}}]{PhysRevLett.129.027002}%
  \BibitemOpen
  \bibfield  {author} {\bibinfo {author} {\bibfnamefont {G.}~\bibnamefont
  {Krieger}}, \bibinfo {author} {\bibfnamefont {L.}~\bibnamefont {Martinelli}},
  \bibinfo {author} {\bibfnamefont {S.}~\bibnamefont {Zeng}}, \bibinfo {author}
  {\bibfnamefont {L.~E.}\ \bibnamefont {Chow}}, \bibinfo {author}
  {\bibfnamefont {K.}~\bibnamefont {Kummer}}, \bibinfo {author} {\bibfnamefont
  {R.}~\bibnamefont {Arpaia}}, \bibinfo {author} {\bibfnamefont
  {M.}~\bibnamefont {Moretti~Sala}}, \bibinfo {author} {\bibfnamefont {N.~B.}\
  \bibnamefont {Brookes}}, \bibinfo {author} {\bibfnamefont {A.}~\bibnamefont
  {Ariando}}, \bibinfo {author} {\bibfnamefont {N.}~\bibnamefont {Viart}},
  \bibinfo {author} {\bibfnamefont {M.}~\bibnamefont {Salluzzo}}, \bibinfo
  {author} {\bibfnamefont {G.}~\bibnamefont {Ghiringhelli}}, \ and\ \bibinfo
  {author} {\bibfnamefont {D.}~\bibnamefont {Preziosi}},\ }\href {\doibase
  10.1103/PhysRevLett.129.027002} {\bibfield  {journal} {\bibinfo  {journal}
  {Phys. Rev. Lett.}\ }\textbf {\bibinfo {volume} {129}},\ \bibinfo {pages}
  {027002} (\bibinfo {year} {2022})}\BibitemShut {NoStop}%
\bibitem [{\citenamefont {Peng}\ \emph {et~al.}(2021)\citenamefont {Peng},
  \citenamefont {Jiang}, \citenamefont {Moritz}, \citenamefont {Devereaux},\
  and\ \citenamefont {Jia}}]{peng2021superconductivity}%
  \BibitemOpen
  \bibfield  {author} {\bibinfo {author} {\bibfnamefont {C.}~\bibnamefont
  {Peng}}, \bibinfo {author} {\bibfnamefont {H.-C.}\ \bibnamefont {Jiang}},
  \bibinfo {author} {\bibfnamefont {B.}~\bibnamefont {Moritz}}, \bibinfo
  {author} {\bibfnamefont {T.~P.}\ \bibnamefont {Devereaux}}, \ and\ \bibinfo
  {author} {\bibfnamefont {C.}~\bibnamefont {Jia}},\ }\href@noop {} {}
  (\bibinfo {year} {2021}),\ \Eprint {http://arxiv.org/abs/2110.07593}
  {arXiv:2110.07593} \BibitemShut {NoStop}%
\bibitem [{\citenamefont {Chen}\ \emph {et~al.}(2022)\citenamefont {Chen},
  \citenamefont {feng Yang},\ and\ \citenamefont {Zhang}}]{arXiv:2204.12208}%
  \BibitemOpen
  \bibfield  {author} {\bibinfo {author} {\bibfnamefont {H.}~\bibnamefont
  {Chen}}, \bibinfo {author} {\bibfnamefont {Y.}~\bibnamefont {feng Yang}}, \
  and\ \bibinfo {author} {\bibfnamefont {G.-M.}\ \bibnamefont {Zhang}},\
  }\href@noop {} {} (\bibinfo {year} {2022}),\ \Eprint
  {http://arxiv.org/abs/2204.12208} {arXiv:2204.12208} \BibitemShut {NoStop}%
\bibitem [{\citenamefont {Shen}\ \emph {et~al.}(2022)\citenamefont {Shen},
  \citenamefont {Qin},\ and\ \citenamefont {Zhang}}]{arXiv:2207.00266}%
  \BibitemOpen
  \bibfield  {author} {\bibinfo {author} {\bibfnamefont {Y.}~\bibnamefont
  {Shen}}, \bibinfo {author} {\bibfnamefont {M.}~\bibnamefont {Qin}}, \ and\
  \bibinfo {author} {\bibfnamefont {G.-M.}\ \bibnamefont {Zhang}},\ }\href@noop
  {} {} (\bibinfo {year} {2022}),\ \Eprint {http://arxiv.org/abs/2207.00266}
  {arXiv:2207.00266} \BibitemShut {NoStop}%
\bibitem [{\citenamefont {Anisimov}\ \emph {et~al.}(1999)\citenamefont
  {Anisimov}, \citenamefont {Bukhvalov},\ and\ \citenamefont
  {Rice}}]{PhysRevB.59.7901}%
  \BibitemOpen
  \bibfield  {author} {\bibinfo {author} {\bibfnamefont {V.~I.}\ \bibnamefont
  {Anisimov}}, \bibinfo {author} {\bibfnamefont {D.}~\bibnamefont {Bukhvalov}},
  \ and\ \bibinfo {author} {\bibfnamefont {T.~M.}\ \bibnamefont {Rice}},\
  }\href {\doibase 10.1103/PhysRevB.59.7901} {\bibfield  {journal} {\bibinfo
  {journal} {Phys. Rev. B}\ }\textbf {\bibinfo {volume} {59}},\ \bibinfo
  {pages} {7901} (\bibinfo {year} {1999})}\BibitemShut {NoStop}%
\bibitem [{\citenamefont {Lee}\ and\ \citenamefont
  {Pickett}(2004)}]{PhysRevB.70.165109}%
  \BibitemOpen
  \bibfield  {author} {\bibinfo {author} {\bibfnamefont {K.-W.}\ \bibnamefont
  {Lee}}\ and\ \bibinfo {author} {\bibfnamefont {W.~E.}\ \bibnamefont
  {Pickett}},\ }\href {\doibase 10.1103/PhysRevB.70.165109} {\bibfield
  {journal} {\bibinfo  {journal} {Phys. Rev. B}\ }\textbf {\bibinfo {volume}
  {70}},\ \bibinfo {pages} {165109} (\bibinfo {year} {2004})}\BibitemShut
  {NoStop}%
\bibitem [{\citenamefont {Botana}\ and\ \citenamefont
  {Norman}(2020)}]{PhysRevX.10.011024}%
  \BibitemOpen
  \bibfield  {author} {\bibinfo {author} {\bibfnamefont {A.~S.}\ \bibnamefont
  {Botana}}\ and\ \bibinfo {author} {\bibfnamefont {M.~R.}\ \bibnamefont
  {Norman}},\ }\href {\doibase 10.1103/PhysRevX.10.011024} {\bibfield
  {journal} {\bibinfo  {journal} {Phys. Rev. X}\ }\textbf {\bibinfo {volume}
  {10}},\ \bibinfo {pages} {011024} (\bibinfo {year} {2020})}\BibitemShut
  {NoStop}%
\bibitem [{\citenamefont {Been}\ \emph {et~al.}(2021)\citenamefont {Been},
  \citenamefont {Lee}, \citenamefont {Hwang}, \citenamefont {Cui},
  \citenamefont {Zaanen}, \citenamefont {Devereaux}, \citenamefont {Moritz},\
  and\ \citenamefont {Jia}}]{PhysRevX.11.011050}%
  \BibitemOpen
  \bibfield  {author} {\bibinfo {author} {\bibfnamefont {E.}~\bibnamefont
  {Been}}, \bibinfo {author} {\bibfnamefont {W.-S.}\ \bibnamefont {Lee}},
  \bibinfo {author} {\bibfnamefont {H.~Y.}\ \bibnamefont {Hwang}}, \bibinfo
  {author} {\bibfnamefont {Y.}~\bibnamefont {Cui}}, \bibinfo {author}
  {\bibfnamefont {J.}~\bibnamefont {Zaanen}}, \bibinfo {author} {\bibfnamefont
  {T.}~\bibnamefont {Devereaux}}, \bibinfo {author} {\bibfnamefont
  {B.}~\bibnamefont {Moritz}}, \ and\ \bibinfo {author} {\bibfnamefont
  {C.}~\bibnamefont {Jia}},\ }\href {\doibase 10.1103/PhysRevX.11.011050}
  {\bibfield  {journal} {\bibinfo  {journal} {Phys. Rev. X}\ }\textbf {\bibinfo
  {volume} {11}},\ \bibinfo {pages} {011050} (\bibinfo {year}
  {2021})}\BibitemShut {NoStop}%
\bibitem [{\citenamefont {Hepting}\ \emph {et~al.}(2020)\citenamefont
  {Hepting}, \citenamefont {Li}, \citenamefont {Jia}, \citenamefont {Lu},
  \citenamefont {Paris}, \citenamefont {Tseng}, \citenamefont {Feng},
  \citenamefont {Osada}, \citenamefont {Been}, \citenamefont {Hikita},
  \citenamefont {Chuang}, \citenamefont {Hussain}, \citenamefont {Zhou},
  \citenamefont {Nag}, \citenamefont {Garcia-Fernandez}, \citenamefont {Rossi},
  \citenamefont {Huang}, \citenamefont {Huang}, \citenamefont {Shen},
  \citenamefont {Schmitt}, \citenamefont {Hwang}, \citenamefont {Moritz},
  \citenamefont {Zaanen}, \citenamefont {Devereaux},\ and\ \citenamefont
  {Lee}}]{Hepting2020}%
  \BibitemOpen
  \bibfield  {author} {\bibinfo {author} {\bibfnamefont {M.}~\bibnamefont
  {Hepting}}, \bibinfo {author} {\bibfnamefont {D.}~\bibnamefont {Li}},
  \bibinfo {author} {\bibfnamefont {C.~J.}\ \bibnamefont {Jia}}, \bibinfo
  {author} {\bibfnamefont {H.}~\bibnamefont {Lu}}, \bibinfo {author}
  {\bibfnamefont {E.}~\bibnamefont {Paris}}, \bibinfo {author} {\bibfnamefont
  {Y.}~\bibnamefont {Tseng}}, \bibinfo {author} {\bibfnamefont
  {X.}~\bibnamefont {Feng}}, \bibinfo {author} {\bibfnamefont {M.}~\bibnamefont
  {Osada}}, \bibinfo {author} {\bibfnamefont {E.}~\bibnamefont {Been}},
  \bibinfo {author} {\bibfnamefont {Y.}~\bibnamefont {Hikita}}, \bibinfo
  {author} {\bibfnamefont {Y.-D.}\ \bibnamefont {Chuang}}, \bibinfo {author}
  {\bibfnamefont {Z.}~\bibnamefont {Hussain}}, \bibinfo {author} {\bibfnamefont
  {K.~J.}\ \bibnamefont {Zhou}}, \bibinfo {author} {\bibfnamefont
  {A.}~\bibnamefont {Nag}}, \bibinfo {author} {\bibfnamefont {M.}~\bibnamefont
  {Garcia-Fernandez}}, \bibinfo {author} {\bibfnamefont {M.}~\bibnamefont
  {Rossi}}, \bibinfo {author} {\bibfnamefont {H.~Y.}\ \bibnamefont {Huang}},
  \bibinfo {author} {\bibfnamefont {D.~J.}\ \bibnamefont {Huang}}, \bibinfo
  {author} {\bibfnamefont {Z.~X.}\ \bibnamefont {Shen}}, \bibinfo {author}
  {\bibfnamefont {T.}~\bibnamefont {Schmitt}}, \bibinfo {author} {\bibfnamefont
  {H.~Y.}\ \bibnamefont {Hwang}}, \bibinfo {author} {\bibfnamefont
  {B.}~\bibnamefont {Moritz}}, \bibinfo {author} {\bibfnamefont
  {J.}~\bibnamefont {Zaanen}}, \bibinfo {author} {\bibfnamefont {T.~P.}\
  \bibnamefont {Devereaux}}, \ and\ \bibinfo {author} {\bibfnamefont {W.~S.}\
  \bibnamefont {Lee}},\ }\href {\doibase 10.1038/s41563-019-0585-z} {\bibfield
  {journal} {\bibinfo  {journal} {Nature Materials}\ }\textbf {\bibinfo
  {volume} {19}},\ \bibinfo {pages} {381} (\bibinfo {year} {2020})}\BibitemShut
  {NoStop}%
\bibitem [{\citenamefont {Jiang}\ \emph {et~al.}(2020)\citenamefont {Jiang},
  \citenamefont {Berciu},\ and\ \citenamefont
  {Sawatzky}}]{PhysRevLett.124.207004}%
  \BibitemOpen
  \bibfield  {author} {\bibinfo {author} {\bibfnamefont {M.}~\bibnamefont
  {Jiang}}, \bibinfo {author} {\bibfnamefont {M.}~\bibnamefont {Berciu}}, \
  and\ \bibinfo {author} {\bibfnamefont {G.~A.}\ \bibnamefont {Sawatzky}},\
  }\href {\doibase 10.1103/PhysRevLett.124.207004} {\bibfield  {journal}
  {\bibinfo  {journal} {Phys. Rev. Lett.}\ }\textbf {\bibinfo {volume} {124}},\
  \bibinfo {pages} {207004} (\bibinfo {year} {2020})}\BibitemShut {NoStop}%
\bibitem [{\citenamefont {Nomura}\ \emph {et~al.}(2019)\citenamefont {Nomura},
  \citenamefont {Hirayama}, \citenamefont {Tadano}, \citenamefont {Yoshimoto},
  \citenamefont {Nakamura},\ and\ \citenamefont {Arita}}]{PhysRevB.100.205138}%
  \BibitemOpen
  \bibfield  {author} {\bibinfo {author} {\bibfnamefont {Y.}~\bibnamefont
  {Nomura}}, \bibinfo {author} {\bibfnamefont {M.}~\bibnamefont {Hirayama}},
  \bibinfo {author} {\bibfnamefont {T.}~\bibnamefont {Tadano}}, \bibinfo
  {author} {\bibfnamefont {Y.}~\bibnamefont {Yoshimoto}}, \bibinfo {author}
  {\bibfnamefont {K.}~\bibnamefont {Nakamura}}, \ and\ \bibinfo {author}
  {\bibfnamefont {R.}~\bibnamefont {Arita}},\ }\href {\doibase
  10.1103/PhysRevB.100.205138} {\bibfield  {journal} {\bibinfo  {journal}
  {Phys. Rev. B}\ }\textbf {\bibinfo {volume} {100}},\ \bibinfo {pages}
  {205138} (\bibinfo {year} {2019})}\BibitemShut {NoStop}%
\bibitem [{\citenamefont {Hu}\ and\ \citenamefont
  {Wu}(2019)}]{PhysRevResearch.1.032046}%
  \BibitemOpen
  \bibfield  {author} {\bibinfo {author} {\bibfnamefont {L.-H.}\ \bibnamefont
  {Hu}}\ and\ \bibinfo {author} {\bibfnamefont {C.}~\bibnamefont {Wu}},\ }\href
  {\doibase 10.1103/PhysRevResearch.1.032046} {\bibfield  {journal} {\bibinfo
  {journal} {Phys. Rev. Research}\ }\textbf {\bibinfo {volume} {1}},\ \bibinfo
  {pages} {032046} (\bibinfo {year} {2019})}\BibitemShut {NoStop}%
\bibitem [{\citenamefont {Karp}\ \emph {et~al.}(2020)\citenamefont {Karp},
  \citenamefont {Botana}, \citenamefont {Norman}, \citenamefont {Park},
  \citenamefont {Zingl},\ and\ \citenamefont {Millis}}]{PhysRevX.10.021061}%
  \BibitemOpen
  \bibfield  {author} {\bibinfo {author} {\bibfnamefont {J.}~\bibnamefont
  {Karp}}, \bibinfo {author} {\bibfnamefont {A.~S.}\ \bibnamefont {Botana}},
  \bibinfo {author} {\bibfnamefont {M.~R.}\ \bibnamefont {Norman}}, \bibinfo
  {author} {\bibfnamefont {H.}~\bibnamefont {Park}}, \bibinfo {author}
  {\bibfnamefont {M.}~\bibnamefont {Zingl}}, \ and\ \bibinfo {author}
  {\bibfnamefont {A.}~\bibnamefont {Millis}},\ }\href {\doibase
  10.1103/PhysRevX.10.021061} {\bibfield  {journal} {\bibinfo  {journal} {Phys.
  Rev. X}\ }\textbf {\bibinfo {volume} {10}},\ \bibinfo {pages} {021061}
  (\bibinfo {year} {2020})}\BibitemShut {NoStop}%
\bibitem [{\citenamefont {Lechermann}(2020)}]{PhysRevX.10.041002}%
  \BibitemOpen
  \bibfield  {author} {\bibinfo {author} {\bibfnamefont {F.}~\bibnamefont
  {Lechermann}},\ }\href {\doibase 10.1103/PhysRevX.10.041002} {\bibfield
  {journal} {\bibinfo  {journal} {Phys. Rev. X}\ }\textbf {\bibinfo {volume}
  {10}},\ \bibinfo {pages} {041002} (\bibinfo {year} {2020})}\BibitemShut
  {NoStop}%
\bibitem [{\citenamefont {Blankenbecler}\ \emph {et~al.}(1981)\citenamefont
  {Blankenbecler}, \citenamefont {Scalapino},\ and\ \citenamefont
  {Sugar}}]{PhysRevD.24.2278}%
  \BibitemOpen
  \bibfield  {author} {\bibinfo {author} {\bibfnamefont {R.}~\bibnamefont
  {Blankenbecler}}, \bibinfo {author} {\bibfnamefont {D.~J.}\ \bibnamefont
  {Scalapino}}, \ and\ \bibinfo {author} {\bibfnamefont {R.~L.}\ \bibnamefont
  {Sugar}},\ }\href {\doibase 10.1103/PhysRevD.24.2278} {\bibfield  {journal}
  {\bibinfo  {journal} {Phys. Rev. D}\ }\textbf {\bibinfo {volume} {24}},\
  \bibinfo {pages} {2278} (\bibinfo {year} {1981})}\BibitemShut {NoStop}%
\bibitem [{\citenamefont {Ma}\ \emph {et~al.}(2010)\citenamefont {Ma},
  \citenamefont {Hu}, \citenamefont {Huang},\ and\ \citenamefont
  {Lin}}]{doi:10.1063/1.3485059}%
  \BibitemOpen
  \bibfield  {author} {\bibinfo {author} {\bibfnamefont {T.}~\bibnamefont
  {Ma}}, \bibinfo {author} {\bibfnamefont {F.}~\bibnamefont {Hu}}, \bibinfo
  {author} {\bibfnamefont {Z.}~\bibnamefont {Huang}}, \ and\ \bibinfo {author}
  {\bibfnamefont {H.-Q.}\ \bibnamefont {Lin}},\ }\href {\doibase
  10.1063/1.3485059} {\bibfield  {journal} {\bibinfo  {journal} {Applied
  Physics Letters}\ }\textbf {\bibinfo {volume} {97}},\ \bibinfo {pages}
  {112504} (\bibinfo {year} {2010})}\BibitemShut {NoStop}%
\bibitem [{\citenamefont {Ma}\ \emph {et~al.}(2013)\citenamefont {Ma},
  \citenamefont {Lin},\ and\ \citenamefont {Hu}}]{PhysRevLett.110.107002}%
  \BibitemOpen
  \bibfield  {author} {\bibinfo {author} {\bibfnamefont {T.}~\bibnamefont
  {Ma}}, \bibinfo {author} {\bibfnamefont {H.-Q.}\ \bibnamefont {Lin}}, \ and\
  \bibinfo {author} {\bibfnamefont {J.}~\bibnamefont {Hu}},\ }\href {\doibase
  10.1103/PhysRevLett.110.107002} {\bibfield  {journal} {\bibinfo  {journal}
  {Phys. Rev. Lett.}\ }\textbf {\bibinfo {volume} {110}},\ \bibinfo {pages}
  {107002} (\bibinfo {year} {2013})}\BibitemShut {NoStop}%
\bibitem [{\citenamefont {White}\ \emph {et~al.}(1989)\citenamefont {White},
  \citenamefont {Scalapino}, \citenamefont {Sugar}, \citenamefont {Loh},
  \citenamefont {Gubernatis},\ and\ \citenamefont
  {Scalettar}}]{PhysRevB.40.506}%
  \BibitemOpen
  \bibfield  {author} {\bibinfo {author} {\bibfnamefont {S.~R.}\ \bibnamefont
  {White}}, \bibinfo {author} {\bibfnamefont {D.~J.}\ \bibnamefont
  {Scalapino}}, \bibinfo {author} {\bibfnamefont {R.~L.}\ \bibnamefont
  {Sugar}}, \bibinfo {author} {\bibfnamefont {E.~Y.}\ \bibnamefont {Loh}},
  \bibinfo {author} {\bibfnamefont {J.~E.}\ \bibnamefont {Gubernatis}}, \ and\
  \bibinfo {author} {\bibfnamefont {R.~T.}\ \bibnamefont {Scalettar}},\ }\href
  {\doibase 10.1103/PhysRevB.40.506} {\bibfield  {journal} {\bibinfo  {journal}
  {Phys. Rev. B}\ }\textbf {\bibinfo {volume} {40}},\ \bibinfo {pages} {506}
  (\bibinfo {year} {1989})}\BibitemShut {NoStop}%
\bibitem [{\citenamefont {Wang}\ \emph {et~al.}(2014)\citenamefont {Wang},
  \citenamefont {Corboz},\ and\ \citenamefont {Troyer}}]{Wang_2014}%
  \BibitemOpen
  \bibfield  {author} {\bibinfo {author} {\bibfnamefont {L.}~\bibnamefont
  {Wang}}, \bibinfo {author} {\bibfnamefont {P.}~\bibnamefont {Corboz}}, \ and\
  \bibinfo {author} {\bibfnamefont {M.}~\bibnamefont {Troyer}},\ }\href
  {\doibase 10.1088/1367-2630/16/10/103008} {\bibfield  {journal} {\bibinfo
  {journal} {New Journal of Physics}\ }\textbf {\bibinfo {volume} {16}},\
  \bibinfo {pages} {103008} (\bibinfo {year} {2014})}\BibitemShut {NoStop}%
\bibitem [{\citenamefont {Li}\ \emph {et~al.}(2015)\citenamefont {Li},
  \citenamefont {Jiang},\ and\ \citenamefont {Yao}}]{PhysRevB.91.241117}%
  \BibitemOpen
  \bibfield  {author} {\bibinfo {author} {\bibfnamefont {Z.-X.}\ \bibnamefont
  {Li}}, \bibinfo {author} {\bibfnamefont {Y.-F.}\ \bibnamefont {Jiang}}, \
  and\ \bibinfo {author} {\bibfnamefont {H.}~\bibnamefont {Yao}},\ }\href
  {\doibase 10.1103/PhysRevB.91.241117} {\bibfield  {journal} {\bibinfo
  {journal} {Phys. Rev. B}\ }\textbf {\bibinfo {volume} {91}},\ \bibinfo
  {pages} {241117} (\bibinfo {year} {2015})}\BibitemShut {NoStop}%
\bibitem [{\citenamefont {Zhang}\ \emph {et~al.}(2019)\citenamefont {Zhang},
  \citenamefont {Chiu}, \citenamefont {Costa}, \citenamefont {Batrouni},\ and\
  \citenamefont {Scalettar}}]{PhysRevLett.122.077602}%
  \BibitemOpen
  \bibfield  {author} {\bibinfo {author} {\bibfnamefont {Y.-X.}\ \bibnamefont
  {Zhang}}, \bibinfo {author} {\bibfnamefont {W.-T.}\ \bibnamefont {Chiu}},
  \bibinfo {author} {\bibfnamefont {N.~C.}\ \bibnamefont {Costa}}, \bibinfo
  {author} {\bibfnamefont {G.~G.}\ \bibnamefont {Batrouni}}, \ and\ \bibinfo
  {author} {\bibfnamefont {R.~T.}\ \bibnamefont {Scalettar}},\ }\href {\doibase
  10.1103/PhysRevLett.122.077602} {\bibfield  {journal} {\bibinfo  {journal}
  {Phys. Rev. Lett.}\ }\textbf {\bibinfo {volume} {122}},\ \bibinfo {pages}
  {077602} (\bibinfo {year} {2019})}\BibitemShut {NoStop}%
\bibitem [{\citenamefont {Karp}\ \emph {et~al.}(2022)\citenamefont {Karp},
  \citenamefont {Hampel},\ and\ \citenamefont {Millis}}]{PhysRevB.105.205131}%
  \BibitemOpen
  \bibfield  {author} {\bibinfo {author} {\bibfnamefont {J.}~\bibnamefont
  {Karp}}, \bibinfo {author} {\bibfnamefont {A.}~\bibnamefont {Hampel}}, \ and\
  \bibinfo {author} {\bibfnamefont {A.~J.}\ \bibnamefont {Millis}},\ }\href
  {\doibase 10.1103/PhysRevB.105.205131} {\bibfield  {journal} {\bibinfo
  {journal} {Phys. Rev. B}\ }\textbf {\bibinfo {volume} {105}},\ \bibinfo
  {pages} {205131} (\bibinfo {year} {2022})}\BibitemShut {NoStop}%
\bibitem [{\citenamefont {Klett}\ \emph {et~al.}(2022)\citenamefont {Klett},
  \citenamefont {Hansmann},\ and\ \citenamefont {Schäfer}}]{fphy.2022.834682}%
  \BibitemOpen
  \bibfield  {author} {\bibinfo {author} {\bibfnamefont {M.}~\bibnamefont
  {Klett}}, \bibinfo {author} {\bibfnamefont {P.}~\bibnamefont {Hansmann}}, \
  and\ \bibinfo {author} {\bibfnamefont {T.}~\bibnamefont {Schäfer}},\ }\href
  {https://www.frontiersin.org/article/10.3389/fphy.2022.834682} {\bibfield
  {journal} {\bibinfo  {journal} {Frontiers in Physics}\ }\textbf {\bibinfo
  {volume} {10}} (\bibinfo {year} {2022})}\BibitemShut {NoStop}%
\bibitem [{\citenamefont {Harvey}\ \emph {et~al.}(2022)\citenamefont {Harvey},
  \citenamefont {Wang}, \citenamefont {Fowlie}, \citenamefont {Osada},
  \citenamefont {Lee}, \citenamefont {Lee}, \citenamefont {Li},\ and\
  \citenamefont {Hwang}}]{arxiv.2201.12971}%
  \BibitemOpen
  \bibfield  {author} {\bibinfo {author} {\bibfnamefont {S.~P.}\ \bibnamefont
  {Harvey}}, \bibinfo {author} {\bibfnamefont {B.~Y.}\ \bibnamefont {Wang}},
  \bibinfo {author} {\bibfnamefont {J.}~\bibnamefont {Fowlie}}, \bibinfo
  {author} {\bibfnamefont {M.}~\bibnamefont {Osada}}, \bibinfo {author}
  {\bibfnamefont {K.}~\bibnamefont {Lee}}, \bibinfo {author} {\bibfnamefont
  {Y.}~\bibnamefont {Lee}}, \bibinfo {author} {\bibfnamefont {D.}~\bibnamefont
  {Li}}, \ and\ \bibinfo {author} {\bibfnamefont {H.~Y.}\ \bibnamefont
  {Hwang}},\ }\href@noop {} {} (\bibinfo {year} {2022}),\ \Eprint
  {http://arxiv.org/abs/2201.12971} {arXiv:2201.12971} \BibitemShut {NoStop}%
\bibitem [{\citenamefont {Hayward}\ and\ \citenamefont
  {Rosseinsky}(2003)}]{HAYWARD2003839}%
  \BibitemOpen
  \bibfield  {author} {\bibinfo {author} {\bibfnamefont {M.}~\bibnamefont
  {Hayward}}\ and\ \bibinfo {author} {\bibfnamefont {M.}~\bibnamefont
  {Rosseinsky}},\ }\href {\doibase
  https://doi.org/10.1016/S1293-2558(03)00111-0} {\bibfield  {journal}
  {\bibinfo  {journal} {Solid State Sciences}\ }\textbf {\bibinfo {volume}
  {5}},\ \bibinfo {pages} {839} (\bibinfo {year} {2003})}\BibitemShut {NoStop}%
\bibitem [{\citenamefont {Sui}\ \emph {et~al.}(2022)\citenamefont {Sui},
  \citenamefont {Wang}, \citenamefont {Ding}, \citenamefont {Zhou},
  \citenamefont {Qiao}, \citenamefont {Lin},\ and\ \citenamefont
  {Huang}}]{arXiv:2202.11904}%
  \BibitemOpen
  \bibfield  {author} {\bibinfo {author} {\bibfnamefont {X.}~\bibnamefont
  {Sui}}, \bibinfo {author} {\bibfnamefont {J.}~\bibnamefont {Wang}}, \bibinfo
  {author} {\bibfnamefont {X.}~\bibnamefont {Ding}}, \bibinfo {author}
  {\bibfnamefont {K.-J.}\ \bibnamefont {Zhou}}, \bibinfo {author}
  {\bibfnamefont {L.}~\bibnamefont {Qiao}}, \bibinfo {author} {\bibfnamefont
  {H.}~\bibnamefont {Lin}}, \ and\ \bibinfo {author} {\bibfnamefont
  {B.}~\bibnamefont {Huang}},\ }\href@noop {} {} (\bibinfo {year} {2022}),\
  \Eprint {http://arxiv.org/abs/2202.11904} {arXiv:2202.11904} \BibitemShut
  {NoStop}%
\bibitem [{\citenamefont {Chen}\ \emph {et~al.}()\citenamefont {Chen},
  \citenamefont {Ma}, \citenamefont {Sui}, \citenamefont {Liang}, \citenamefont
  {Huang},\ and\ \citenamefont {Ma}}]{key1}%
  \BibitemOpen
  \bibfield  {author} {\bibinfo {author} {\bibfnamefont {C.}~\bibnamefont
  {Chen}}, \bibinfo {author} {\bibfnamefont {R.}~\bibnamefont {Ma}}, \bibinfo
  {author} {\bibfnamefont {X.}~\bibnamefont {Sui}}, \bibinfo {author}
  {\bibfnamefont {Y.}~\bibnamefont {Liang}}, \bibinfo {author} {\bibfnamefont
  {B.}~\bibnamefont {Huang}}, \ and\ \bibinfo {author} {\bibfnamefont
  {T.}~\bibnamefont {Ma}},\ }\href@noop {} {}\bibinfo {note}
  {(unpublished)}\BibitemShut {NoStop}%
\bibitem [{\citenamefont {Mostofi}\ \emph {et~al.}(2008)\citenamefont
  {Mostofi}, \citenamefont {Yates}, \citenamefont {Lee}, \citenamefont {Souza},
  \citenamefont {Vanderbilt},\ and\ \citenamefont {Marzari}}]{MOSTOFI2008685}%
  \BibitemOpen
  \bibfield  {author} {\bibinfo {author} {\bibfnamefont {A.~A.}\ \bibnamefont
  {Mostofi}}, \bibinfo {author} {\bibfnamefont {J.~R.}\ \bibnamefont {Yates}},
  \bibinfo {author} {\bibfnamefont {Y.-S.}\ \bibnamefont {Lee}}, \bibinfo
  {author} {\bibfnamefont {I.}~\bibnamefont {Souza}}, \bibinfo {author}
  {\bibfnamefont {D.}~\bibnamefont {Vanderbilt}}, \ and\ \bibinfo {author}
  {\bibfnamefont {N.}~\bibnamefont {Marzari}},\ }\href {\doibase
  https://doi.org/10.1016/j.cpc.2007.11.016} {\bibfield  {journal} {\bibinfo
  {journal} {Computer Physics Communications}\ }\textbf {\bibinfo {volume}
  {178}},\ \bibinfo {pages} {685} (\bibinfo {year} {2008})}\BibitemShut
  {NoStop}%
\bibitem [{\citenamefont {Sprau}\ \emph {et~al.}(2017)\citenamefont {Sprau},
  \citenamefont {Kostin}, \citenamefont {Kreisel}, \citenamefont {Böhmer},
  \citenamefont {Taufour}, \citenamefont {Canfield}, \citenamefont {Mukherjee},
  \citenamefont {Hirschfeld}, \citenamefont {Andersen},\ and\ \citenamefont
  {Davis}}]{science.aal1575}%
  \BibitemOpen
  \bibfield  {author} {\bibinfo {author} {\bibfnamefont {P.~O.}\ \bibnamefont
  {Sprau}}, \bibinfo {author} {\bibfnamefont {A.}~\bibnamefont {Kostin}},
  \bibinfo {author} {\bibfnamefont {A.}~\bibnamefont {Kreisel}}, \bibinfo
  {author} {\bibfnamefont {A.~E.}\ \bibnamefont {Böhmer}}, \bibinfo {author}
  {\bibfnamefont {V.}~\bibnamefont {Taufour}}, \bibinfo {author} {\bibfnamefont
  {P.~C.}\ \bibnamefont {Canfield}}, \bibinfo {author} {\bibfnamefont
  {S.}~\bibnamefont {Mukherjee}}, \bibinfo {author} {\bibfnamefont {P.~J.}\
  \bibnamefont {Hirschfeld}}, \bibinfo {author} {\bibfnamefont {B.~M.}\
  \bibnamefont {Andersen}}, \ and\ \bibinfo {author} {\bibfnamefont {J.~C.~S.}\
  \bibnamefont {Davis}},\ }\href {\doibase 10.1126/science.aal1575} {\bibfield
  {journal} {\bibinfo  {journal} {Science}\ }\textbf {\bibinfo {volume}
  {357}},\ \bibinfo {pages} {75} (\bibinfo {year} {2017})}\BibitemShut
  {NoStop}%
\bibitem [{\citenamefont {B\'eri}\ \emph {et~al.}(2009)\citenamefont {B\'eri},
  \citenamefont {Kupferschmidt}, \citenamefont {Beenakker},\ and\ \citenamefont
  {Brouwer}}]{PhysRevB.79.024517}%
  \BibitemOpen
  \bibfield  {author} {\bibinfo {author} {\bibfnamefont {B.}~\bibnamefont
  {B\'eri}}, \bibinfo {author} {\bibfnamefont {J.~N.}\ \bibnamefont
  {Kupferschmidt}}, \bibinfo {author} {\bibfnamefont {C.~W.~J.}\ \bibnamefont
  {Beenakker}}, \ and\ \bibinfo {author} {\bibfnamefont {P.~W.}\ \bibnamefont
  {Brouwer}},\ }\href {\doibase 10.1103/PhysRevB.79.024517} {\bibfield
  {journal} {\bibinfo  {journal} {Phys. Rev. B}\ }\textbf {\bibinfo {volume}
  {79}},\ \bibinfo {pages} {024517} (\bibinfo {year} {2009})}\BibitemShut
  {NoStop}%
\bibitem [{\citenamefont {Ai}\ \emph {et~al.}(2019)\citenamefont {Ai},
  \citenamefont {Gao}, \citenamefont {Liu}, \citenamefont {Zhang},
  \citenamefont {Li}, \citenamefont {Huang}, \citenamefont {Song},
  \citenamefont {Yan}, \citenamefont {Zhao}, \citenamefont {Liu}, \citenamefont
  {Gu}, \citenamefont {Zhang}, \citenamefont {Yang}, \citenamefont {Peng},
  \citenamefont {Xu},\ and\ \citenamefont {Zhou}}]{Ai_2019}%
  \BibitemOpen
  \bibfield  {author} {\bibinfo {author} {\bibfnamefont {P.}~\bibnamefont
  {Ai}}, \bibinfo {author} {\bibfnamefont {Q.}~\bibnamefont {Gao}}, \bibinfo
  {author} {\bibfnamefont {J.}~\bibnamefont {Liu}}, \bibinfo {author}
  {\bibfnamefont {Y.}~\bibnamefont {Zhang}}, \bibinfo {author} {\bibfnamefont
  {C.}~\bibnamefont {Li}}, \bibinfo {author} {\bibfnamefont {J.}~\bibnamefont
  {Huang}}, \bibinfo {author} {\bibfnamefont {C.}~\bibnamefont {Song}},
  \bibinfo {author} {\bibfnamefont {H.}~\bibnamefont {Yan}}, \bibinfo {author}
  {\bibfnamefont {L.}~\bibnamefont {Zhao}}, \bibinfo {author} {\bibfnamefont
  {G.-D.}\ \bibnamefont {Liu}}, \bibinfo {author} {\bibfnamefont {G.-D.}\
  \bibnamefont {Gu}}, \bibinfo {author} {\bibfnamefont {F.-F.}\ \bibnamefont
  {Zhang}}, \bibinfo {author} {\bibfnamefont {F.}~\bibnamefont {Yang}},
  \bibinfo {author} {\bibfnamefont {Q.-J.}\ \bibnamefont {Peng}}, \bibinfo
  {author} {\bibfnamefont {Z.-Y.}\ \bibnamefont {Xu}}, \ and\ \bibinfo {author}
  {\bibfnamefont {X.-J.}\ \bibnamefont {Zhou}},\ }\href {\doibase
  10.1088/0256-307x/36/6/067402} {\bibfield  {journal} {\bibinfo  {journal}
  {Chinese Physics Letters}\ }\textbf {\bibinfo {volume} {36}},\ \bibinfo
  {pages} {067402} (\bibinfo {year} {2019})}\BibitemShut {NoStop}%
\bibitem [{\citenamefont {Santos}(2003)}]{SANTOS2003}%
  \BibitemOpen
  \bibfield  {author} {\bibinfo {author} {\bibfnamefont {R.~R.~d.}\
  \bibnamefont {Santos}},\ }\href
  {http://www.scielo.br/scielo.php?script=sci_arttext&pid=S0103-97332003000100003&nrm=iso}
  {\bibfield  {journal} {\bibinfo  {journal} {{Braz. J. Phys.}}\ }\textbf
  {\bibinfo {volume} {33}},\ \bibinfo {pages} {36 } (\bibinfo {year}
  {2003})}\BibitemShut {NoStop}%
\end{thebibliography}%
\end{document}